\documentclass{emulateapj}
\usepackage{textcomp}

\newcommand{\sm}{$\sigma_{\mu>3}$ }
\newcommand{\smd}{$\sigma_{\mu>3}$}
\newcommand{\zten}{$z \sim 10$ }

\shorttitle{Cosmic Telescopes in Millennium}
\shortauthors{French et al.}

\begin{document}

\title{Characterizing the Best Cosmic Telescopes with the Millennium Simulations}
%\date{ \today}

\author{K. Decker French \altaffilmark{1} \altaffilmark{$\dagger$}, Kenneth C. Wong \altaffilmark{1} \altaffilmark{2} \altaffilmark{3}, Ann I. Zabludoff \altaffilmark{1}, \\ S. Mark Ammons \altaffilmark{4}, Charles R. Keeton \altaffilmark{5}, Raul E. Angulo \altaffilmark{6} }
\altaffiltext{1}{Steward Observatory, University of Arizona, 933 N. Cherry Ave, Tucson, AZ 85721}
\altaffiltext{$\dagger$}{kfrench@as.arizona.edu}
\altaffiltext{2}{Institute of Astronomy and Astrophysics, Academia Sinica (ASIAA), Taipei 10641, Taiwan}
\altaffiltext{3}{EACOA Fellow}
\altaffiltext{4}{Lawrence Livermore National Laboratory, 7000 East Ave, Livermore, CA 94550}
\altaffiltext{5}{Department of Physics and Astronomy, Rutgers University, 136 Frelinghuysen Rd., Piscataway, NJ 08854}
\altaffiltext{6}{Centro de Estudios de F\'isica del Cosmos de Arag\'on, Plaza San Juan 1, Planta-2, 44001, Teruel, Spain}

\begin{abstract}
Certain configurations of massive structures projected along the line of sight maximize the number of detections of gravitationally lensed $z\sim10$ galaxies. We characterize such lines of sight with the \'etendue $\sigma_\mu$, the area in the source plane magnified over some threshold $\mu$. We use the Millennium I and Millennium XXL cosmological simulations to determine the frequency of high $\sigma_\mu$ beams on the sky, their properties, and efficient selection criteria. We define the best beams as having $\sigma_{\mu>3} >2000$ arcsec$^2$, for a \zten source plane, and predict $477 \pm 21$ such beams on the sky. 
The total mass in the beam and $\sigma_{\mu>3}$ are strongly correlated. After controlling for total mass, we find a significant residual correlation between $\sigma_{\mu>3}$ and the number of cluster-scale halos ($>10^{14} M_\sun h^{-1}$) in the beam. 
Beams with $\sigma_{\mu>3} >2000$ arcsec$^2$, which should be best at lensing \zten galaxies, are ten times more likely to contain multiple cluster-scale halos than a single cluster-scale halo. 
Beams containing an Abell 1689-like massive cluster halo often have additional structures along the line of sight, including at least one additional cluster-scale ($M_{200}>10^{14}M_\sun h^{-1}$) halo 28\% of the time.  Selecting beams with multiple, massive structures will lead to enhanced detection of the most distant and intrinsically faint galaxies.

\end{abstract}


\section{Introduction}
Accurate and efficient detection of high redshift galaxies is a current limiting problem in observational cosmology. Reionization is expected to occur at a mean redshift of $z \sim 10.4$ \citep{Komatsu2011}. Observing how the luminosity function of galaxies changes as reionization progresses is essential in understanding how reionization takes place \citep{Robertson2013}.

Recent efforts to observe high redshift galaxies have produced over 100 $z\sim7$ galaxies, and a few \zten candidate galaxies \citep{Bouwens2011,Zheng2012,Coe2012a,Ellis2012}. High redshift galaxies are difficult to observe not only due to their distance, but also their intrinsic faintness. Current methods of detecting high redshift galaxies use blank field studies or known strong gravitational lenses.

Using lensing to magnify high redshift source planes has the drawback of reducing the area surveyed per field, though the depth increases. There have been several studies on this tradeoff \citep{Bouwens2009, Maizy2009}, which conclude that if the effective slope, $\beta$, of the faint end of the luminosity function $\phi(L)$ at high redshift,
\begin{equation}
\beta(z) = - \frac{d(\log{\phi})}{d(\log{L})}
\end{equation}
is greater than 1, the gain in depth from lensing overcomes the loss in area.
This slope increases with redshift, from $\sim0.7$ at lower redshifts to $\sim1$ by $z\sim7$ \citep{Bouwens2012,Oesch2012}, though this measurement is still uncertain at the 20\% level. 

Mass along the line of sight (LOS) towards gravitational lenses boosts lensing power \citep{Wambsganss2005,Hilbert2007,Puchwein2009}. \citet{Wambsganss2005} find that as the source plane gets further away, a higher fraction of gravitational lenses are composed of multiple significant structures. At $z_{source}\sim7.5$, about 50\% of cases studied have a second lens plane with a significant contribution to the total lensing cross section. \citet{Puchwein2009} claim that LOS matter can boost the strong lensing cross section by up to 50\%, and that the effect of uncorrelated structure along the line of sight is more significant in increasing the strong lensing cross section than structure in the same lens plane as the cluster. These analyses are primarily aimed at determining how LOS mass affects already good lenses, but raise the possibility  that the best cosmic telescopes could have a form other than a single massive cluster lens. 

An analysis of the best configurations of dark matter halos for lensing \zten galaxies is done by \citet{Wong2012}.
They examine potential lines of sight with mass configurations optimized to maximize the \'etendue ($\sigma_\mu$), the area in the $z\sim10$ source plane with magnification ($\mu$) above a certain threshold. 
 Although \'etendue is typically used in reference to optical systems, we adopt it here to characterize the effective optical system of 3D configurations of gravitational lenses.
This provides an advantage over traditional lens characterizations, such as the lensing cross section, which generally refers to the cross-sectional area needed to multiply image source galaxies. For our goal of detecting high redshift galaxies, even modest magnifications provide an advantage, so characterizing beams by \sm allows us to balance sampling depth and magnified area to find the most effective gravitational lenses.
\citet{Wong2012} consider lines of sight with multiple structures along the line of sight.
The division of mass among multiple halos has a large effect on \smd.
Due to interactions between lensing potentials, beams with multiple structures in certain idealized configurations may produce lensing effects comparable to single-cluster beams of twice the total mass. Line of sight lensing beams especially benefit when halos are projected closer on the sky, and if the total mass is split into 2-5 halos of about equal mass. 

Knowing the best lens configurations, we can then ask how frequent certain configurations are in the real universe. We want to know the frequency of high \sm beams, which high \sm beam configurations are most likely, how often additional significant lensing mass is projected along the line of sight to a massive cluster, and which observables are most successful at finding the best lensing beams.

In this study, we use several new tools to study the properties of the best lensing beams. By defining a lensing \'etendue, we can better characterize the lenses as observational tools in the search for \zten galaxies. We frame the effect of LOS mass not as a systematic error to be accounted for, but as a way of boosting the total mass and lensing quality of the beam. The best beams will have very large total masses, including very massive cluster-scale halos. 

Drawing halos from simulations requires a large simulation volume in order to sample the high mass tail of the mass function. Here, we use the Millennium XXL simulation \citep{Angulo2012}. At 3 Gpc $h^{-1}$ on a side, it is much larger than simulations used in previous studies of LOS effects in lensing, producing a larger sample of massive lenses. Using simulation halos rather than toy models provides information about the frequency of halo configurations, which we use to calculate the number of beams that we expect to see on the sky with high \'etendue.

We focus on maximizing sources detections at $z\sim10$, as lensing stands to greatly improve number counts and statistics at this redshift.
We present the methods used in drawing sample lines of sight from the simulations and calculating their lensing properties in Section 2. We present the results of these analyses and their implications on observing strategy in Section 3. Discussion of some choices in methods is presented in Section 4. We conclude in Section 5.

%%%%%%%%%%%%%%%%%%%%%%%%%%%%%%%%%%%%%%%%%%%%%%%%%%%%%%%%%%%%%%%%%%%%%%%%%%%%
%%%%%%%%% Methods- section 2 %%%%%%%%%%%%%%%%%%%%%%%%%%%%%%%%%%%%%%%%%%%%%%%
%%%%%%%%%%%%%%%%%%%%%%%%%%%%%%%%%%%%%%%%%%%%%%%%%%%%%%%%%%%%%%%%%%%%%%%%%%%%

\section{Methods}

\subsection{Millennium Simulations}

We use the Millennium (MS, \citet{Springel2005}) and Millennium-XXL (MXXL, \citet{Angulo2012}) simulations to find and study lines of sight with high
lensing \'etendue. These two simulations complement each other: the MXXL simulation
provides us with a large volume and contains very massive ($>10^{15} M_\sun h^{-1}$) halos, whereas the
MS resolves lower mass halos, thus allowing us to test for numerical artifacts, consider a wide dynamic range in halo mass, and to define halo mass cuts to be applied to our samples.

Both simulations were carried out using cosmological parameters consistent with
the first year data of the WMAP satellite \citep{Spergel2001}, namely $\Omega_m
= 0.25$, $\sigma_8=0.9$, $n_s = 1.0$, and $h=0.73$. These parameters are disfavored by
more recent cosmological analyses (e.g. \citet{PlanckCollaboration2013}), though we explicitly
test for the impact of cosmology on our results in Section \ref{cosmology}, finding no significant changes to our qualitative conclusions.

The MS evolved $2160^3$ particles inside a cubical region of 500 Mpc $h^{-1}$ a side,
from $z=120$ to the present day. The MXXL did so for 30 times more particles in a
box of 216 times larger volume. For the choice of cosmological parameters, the
mass of each simulation particle is $8.1\times10^8 M_\sun h^{-1}$ in the MS, and approximately
eight times larger, $6.7\times10^9 M_\sun h^{-1}$, in the MXXL. Without repetition, the simulated
boxes allow one to create all-sky mock light cones up to z=0.06 and z=0.7 in MS and MXXL respectively.
Gravitational forces were computed using memory-efficient versions of the Tree-PM code {\tt Gadget} \citep{Springel2005}, and
were softened on scales below 5 and 10 kpc $h^{-1}$ for the MS and MXXL, respectively. 

In both simulations, dark matter halos were identified using a Friends-of-Friends
algorithm \citep{Davis1985} at 63 discrete output times (roughly equally
spaced by 100 Myr at low redshifts). Within each halo resolved with more than
20 particles, self-bound substructures were located using the SUBFIND algorithm
\citep{Springel2001}. Additionally, a spherical-overdensity mass, $M_{200}$, was
assigned to each main halo. This corresponds to the mass enclosed by a sphere
with mean density equal to 200 times the critical density, centered at the
location of the potential minimum of the parent halo. In the remainder of the
paper we will refer to this quantity as the mass of a halo.

In computing the lensing properties of the simulated cosmic fields, we refrain from using the full particle information to compute the lensing 
properties of the simulated cosmic fields. Instead, we use analytic NFW model halos \citep{Navarro1996}, with a redshift-dependent
concentration-mass relation given by \citet{Zhao2009}, and neglect the
scatter in the relation. Using an analytic profile and average concentration
has the advantage of providing predictions for an {\it average} sky, in which 
the additional scatter due to quantities like ellipticity, dynamical state, and viewing angle
are all averaged out. This is particularly important, as we only have one realization of 
the simulated universe. We discuss the consequences of scatter in concentration and ellipticity, as well as our use of the NFW profile, in Section \ref{haloparams}.

\subsection{Construction of Light Cones}
\label{lc}
We use the data from the MS and MXXL simulation boxes to consider mock observations along various lines of sight. In order to do so, we must construct light cones.

There are several challenges in using the Millennium simulations in considering lines of sight. The length of the simulation box for the MS, 500 Mpc $h^{-1}$, corresponds to a comoving radial distance out to $z\sim0.17$. To simulate light cones at higher redshifts, the simulation is tiled, and light cones are projected through the simulation box multiple times. Strategic lines of sight are used to avoid multiple projections through the same region. For example, a light cone shot from corner to corner would quickly create a kaleidoscope effect of the same cluster at different points in its evolution, and so is not used. Additionally, the simulation data are only available in discrete snapshots in redshift or time. When constructing the light cones, one must decide whether or not to interpolate halo positions between snapshot redshifts, or to move them in space to preserve their redshift. Most light cone constructions in the literature choose the latter option.

The COSMOS cones \citep{Kitzbichler2007} are a dataset within the Millennium simulation, with 24 pencil-beam light cones of 1.4 degrees by 1.4 degrees square. The cones are created by tiling the simulation box, and for each redshift snapshot selecting the galaxies at the correct cosmological distance away. There are 8 cones, starting at 3 corners of the simulation box, chosen to have the farthest distance before repeating a position in the box. 

In MXXL, we construct light cones in the same manner as \citet{Kitzbichler2007}. The size of the simulation corresponds to a comoving distance out to $z\sim1.4$, which allows light cones to be created without tiling the simulation box. We use two different methods in constructing light cones, depending on our science goals. For understanding the frequency of halo configurations we might expect on the sky (Sections 3.1-3.5, 3.7), we select an origin point for an observer, and construct the light cones from that point at z=0. For understanding possible LOS mass around a specific halo (Section 3.6), we construct light cones backward from that point to a sphere of possible origin points surrounding it. This second technique allows access to the full volume of the MXXL simulation at all redshifts in choosing a halo to build the light cone around.

\begin{figure}
\includegraphics[height=2.5in]{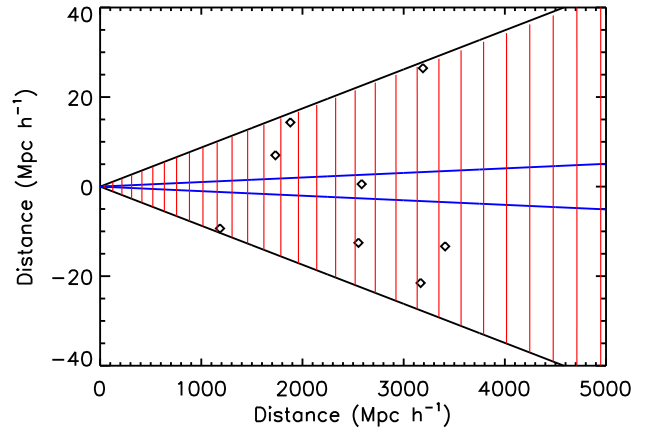}
\caption{Demonstration of light cone construction in MXXL. This figure shows a projection of a 1\textdegree \  by 1\textdegree \  light cone. The inner (blue) lines represent the size of one of our $7^\prime \times 7^\prime$ beams. Vertical lines (red) represent redshift annuli taken out of the simulation box at different redshift snapshots.  Diamonds represent halos in the MXXL simulation falling within this example cone. This method allows us to use a series of redshift snapshots of the simulation box to populate lines of sight with dark matter halos at the appropriate redshifts.}
\label{conepic}
\end{figure}

In the first method, we select an origin point, and objects are binned based on their distance to it. Objects are selected from these bins at simulation redshift steps corresponding to their distance away from the origin. We stitch together annuli from each redshift snapshot of the simulation to create a light cone. This process is illustrated in Figure \ref{conepic}. This figure shows a 1\textdegree \ by 1\textdegree \ light cone from MXXL. Lines representing boundaries in redshift annuli are overplotted, and any halos within these annuli are included in the light cone. Because of the periodic boundary conditions in the simulation box, each of the 8 corners can be selected as an origin point for light cones to simulate a full sky view. In the second method, light cones are constructed backwards from an arbitrary point in the simulation. For a given halo at some position in a redshift snapshot, an annulus of possible origin points exists, and can be used to construct a light cone passing through the selected halo at the proper redshift.

Within each light cone, we select 7$^\prime$ by 7$^\prime$ square beams to study as gravitational lenses. This size is chosen to be consistent with \citet{Wong2012}, and corresponds to a field large enough that massive halos can be separated enough to behave as independent lenses, without any interaction between the lensing potentials. We select the beams out of the light cone using two random numbers simulating right ascension and declination angles. Later, we select only the best lensing beams for further analysis.

As a check on the light cone construction methods, and to ensure results from the two simulations can be combined, we compare the mass function of the halos in our sample of light cones constructed from MXXL to those from the COSMOS cones in the MS halo catalog. We find them to be consistent across epochs from redshifts 0.1-1. 

\subsection{Calculation of $\sigma_\mu$}
\label{smcalc}
To characterize the quality of a beam for maximizing the detection of high redshift sources, we define a quantity that balances the accessible area in the source plane with the lensing magnification at each point. We define the area $\sigma_{\mu>\mu_t}$ in the source plane, where the brightest lensed image of a source has a magnification $\mu$ greater than some threshold value $\mu_t$ \citep{Wong2012}.
This leads to the following definition of $\sigma_{\mu>\mu_t}$ as
\begin{equation}
\sigma_{\mu>\mu_t} = \int_\Omega H(\mathrm{max}(\mu)-\mu_t) d\Omega
\end{equation}
where $H$ is the Heaviside step function, the integral is over $\Omega$, the area in the source plane, $\mathrm{max}(\mu)$ is the magnification of the brightest lensed image of a source at that position in the source plane, and $\mu_t$ is the magnification threshold.  Large areas of intermediate magnifications have been shown to increase the number of detections (Ammons et al. in prep), 
so we choose $\mu_t=3$. The effect of $\mu_t=3$ on the correlation between \sm and number of detections is studied in Section \ref{ndetectsec}. We use a \zten source plane throughout. 
The area in the source plane $\Omega$ is chosen to be large enough to contain the magnified region of the 7$^\prime\times$7$^\prime$ beams. We use a 10$^\prime\times$10$^\prime$ grid in the source plane to accommodate this.

\begin{figure}
\includegraphics[height=3.8in]{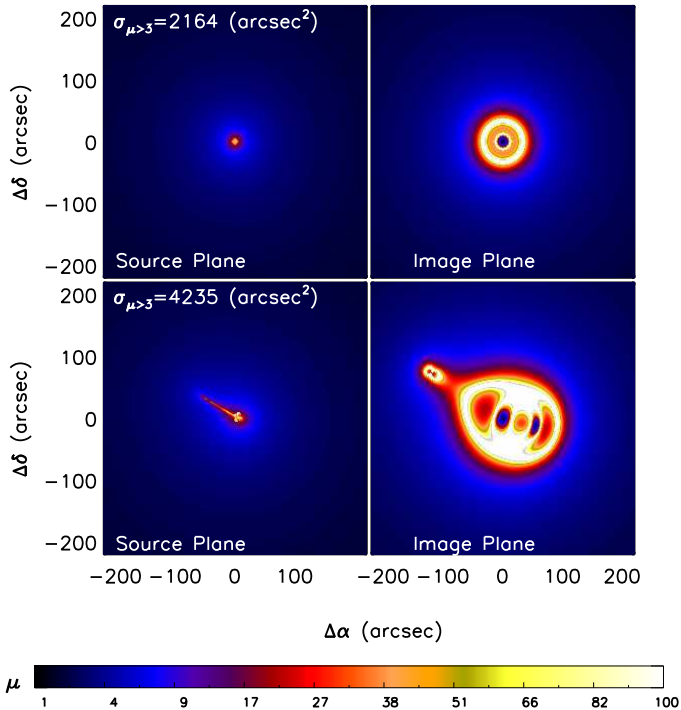}
\caption{Sample magnification maps produced by {\it lensmodel} \citep{Keeton2001}. The panels on the left show a map in the source plane, and the panels on the right show the map in the image plane for the same set of halos. The color scale represents the magnification, $\mu$. The quantity \sm is calculated by integrating the area in the source plane (left hand panels) where the magnification of the brightest image is greater than our threshold value of 3. The beam in the top panels contains a single massive halo of mass $M_{200}=1.9\times 10^{15}M_\sun h^{-1}$. This halo is shown with two additional halos with $M_{200}=1-2\times 10^{14}M_\sun h^{-1}$ found along the line of sight (bottom panels), one projected very close and the other to the upper left. The effect of lensing interaction among the halos can be seen by comparing the top to bottom panels, as the magnified region is larger than if the halos were separated.}
\label{magmap}
\end{figure}

We calculate the magnification map in the source plane using an updated version of  {\it lensmodel} \citep{Keeton2001}. Light from a grid of test sources at $z \sim 10$ is traced through various mass configurations given by beams chosen from the MS and MXXL simulations. The positions, masses, and radii of spherical NFW halos are used to calculate the lensing properties. Each source is mapped to the image plane, and the result is a magnification tensor for each image. We then generate maps of the magnification for a grid of sources in the source and image planes. Example magnification maps in the source plane and image plane are in Figure \ref{magmap}. The two beams shown have the same main halo, with the second showing the effect of additional halos along the line of sight in increasing \smd.

\subsection{Samples and Analysis}
\label{samples}

After constructing the light cones as described in Section \ref{lc}, we select random samples of $7^\prime \times 7^\prime$ beams for further analysis. There is the possibility of overlapping beams in these samples, but this method allows us to accurately determine frequency information. 

We primarily use the MXXL sample for our analyses. We choose eight origin points, each at a corner of the simulation box, from which to construct the light cones. Due to the periodic boundary conditions of the simulation, this is equivalent to generating lightcones that cover the entire sky. The area of the sky can be covered by 3 million 7$^\prime$ by 7$^\prime$ beams. We select 3 million random beams from the simulation. 
Only 30\% of these beams contain at least one halo with a mass greater than $10^{14} M_\sun h^{-1}$ (see Section \ref{masscut}), for a total sample of 1 million beams to study. 

We use the MS sample to justify the $10^{14} M_\sun h^{-1}$ mass cut in the MXXL sample, discussed further in Section 3. For this analysis, we use
10,000 beams, selected from the COSMOS lightcones. Because of the mass cut, ``empty'' beams in the MXXL sample may have many smaller, group-scale halos.

We calculate \sm from the beams in each sample using {\it lensmodel}, as discussed previously.

%%%%%%%%%%%%%%%%%%%%%%%%%%%%%%%%%%%%%%%%%%%%%%%%%%%%%%%%%%%%%%%%%%%%%%%%%%%%
%%%%%%%%% Results- section 3 %%%%%%%%%%%%%%%%%%%%%%%%%%%%%%%%%%%%%%%%%%%%%%%
%%%%%%%%%%%%%%%%%%%%%%%%%%%%%%%%%%%%%%%%%%%%%%%%%%%%%%%%%%%%%%%%%%%%%%%%%%%%

\section{Results and Discussion}
There are two guiding questions we wish to address in this section. The first asks what the best lensing beams look like, and how frequent they are. The second asks which quantities one should use to find those best beams. To answer these questions, we study the dependence of \sm on the beam's total mass, the number of halos, separation in redshift and projected angle, and the sensitivity to cuts in mass and redshift. We then look at the efficiency of selecting on various beam characteristics. 

\subsection{Effect of Total Mass on \'Etendue}

\begin{figure}
\includegraphics[height=2.5in]{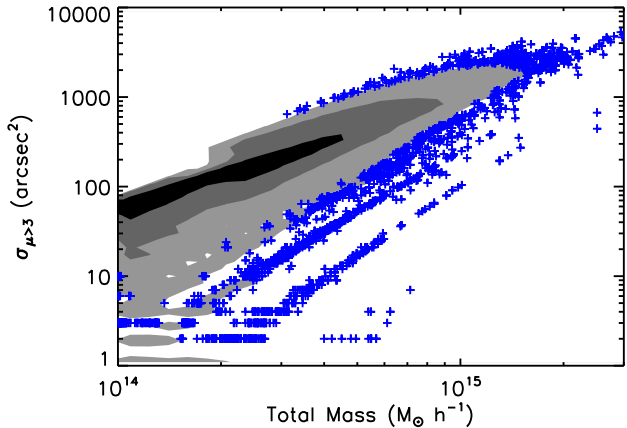}
\caption{\sm dependence on total mass. Total mass is calculated for a $7^\prime \times 7^\prime$ wide beam, for mass within $0<z<1.4$. Contours show the distributions of 68\%, 95\%, and 99.7\% of the beams, with outliers plotted individually (in blue). Total mass is highly correlated with \smd. Discrete stripes of points at low \sm are due to the discrete redshift slices used in constructing the lightcones.}
\label{msigma}
\end{figure}

%\subsection{How massive of a halo is required to make a high \sm beam?}
We know that in general, the more massive a halo, the better lens it will be. Figure \ref{msigma} illustrates this relation between total mass and \sm for beams selected from MXXL, as described in Section \ref{samples}.  Total mass is calculated for a $7^\prime \times 7^\prime$ beam, for mass within $0<z<1.4$, and includes only halos with mass $M_{200}>10^{14}M_\sun h^{-1}$. The relation between total mass and \sm is fit by a power law with index $\alpha=1.36\pm0.02$.
The influence of total mass on \sm is expected to be the most pronounced of any beam quantity, as lensing power depends on the surface mass density.
The scatter in the relation is dominated by the redshift of the halos.
Because of the assumptions we have made about the halo properties, the only difference in \sm between two single-halo beams of the same mass will be due to the difference in redshift, which will enter into the calculation of halo concentration and in the lensing calcuation.  The jump in maximum \sm at $2\times10^{14}M_\sun h^{-1}$ is due to the addition of two-halo beams, which can scatter to higher \sm values, due to the increased number of degrees of freedom. In the following sections, we explore the importance of these quantities, specifically the redshift separation, projected angular separation, and number of halos. But first, we must determine a minimum mass halo to consider in the sample.

\subsection{Minimum Halo Mass to Consider}
\label{masscut}

We cannot include halos down to arbitrarily low masses in MXXL due to storage space limitations (the number of halos increases rapidly with lower mass due to the shape of the halo mass function) and the increased time required to calculate \sm due to additional lens planes. These limits are at a higher mass than the mass resolution of MXXL ($6.7\times10^{9} M_\sun$), and the mass of the smallest resolvable halos ($\approx 10^{11} M_\sun$). We use the Millennium simulation to determine the lowest mass halos that should be considered in studying the highest \sm beams in the MXXL sample.

In order to study the effect of different mass halos on the lensing power, we calculate \sm for each beam in the MS sample several times, each time extending further down in mass. Extending the minimum mass from $10^{15}$ to $10^{14.5}M_\sun h^{-1}$ results in a significant gain in \smd, with the change being over 150\% of the average \smd. Further steps down the mass function to $10^{14}$, $10^{13.5}$, and $10^{13}$ $M_\sun h^{-1}$ result in changes to \sm of less than 20\% of the average. These halos will add to \smd, but will not dominate the effects. Halos with mass less than $10^{14} M_\sun h^{-1}$, even including cases with several halos in a single beam, will not contribute more than a total of 100 arcsec$^2$ to \smd, and halos with mass less than $10^{13} M_\sun h^{-1}$ will not contribute more than a total of 40 arcsec$^2$ to \smd, which is not significant in the \sm$>2000$ arcsec$^2$ range studied here. In our analysis of MXXL, we only consider halos with mass $M_{200}>10^{14}M_\sun h^{-1}$.

\subsection{Frequency of Best Lensing Beams}
 
The best lensing beams in toy models and observations of beams with high total LRG luminosity \citep{Wong2012, S.MarkAmmonsKennethC.WongCharlesR.Keeton2013, Wong2013} are in the range of \sm$=2000-10000$ arcsec$^2$, also using a source plane at $z\sim10$ and a magnification threshold of $\mu_t=3$.  
We define a simulated beam as being a good lensing beam if it has \sm $>1000$ arcsec$^2$, and among the best lensing beams if it has \sm $>2000$ arcsec$^2$, again using a threshold $\mu_t=3$, and a source plane at $z\sim10$. 

\begin{figure}
\includegraphics[width=3.4in]{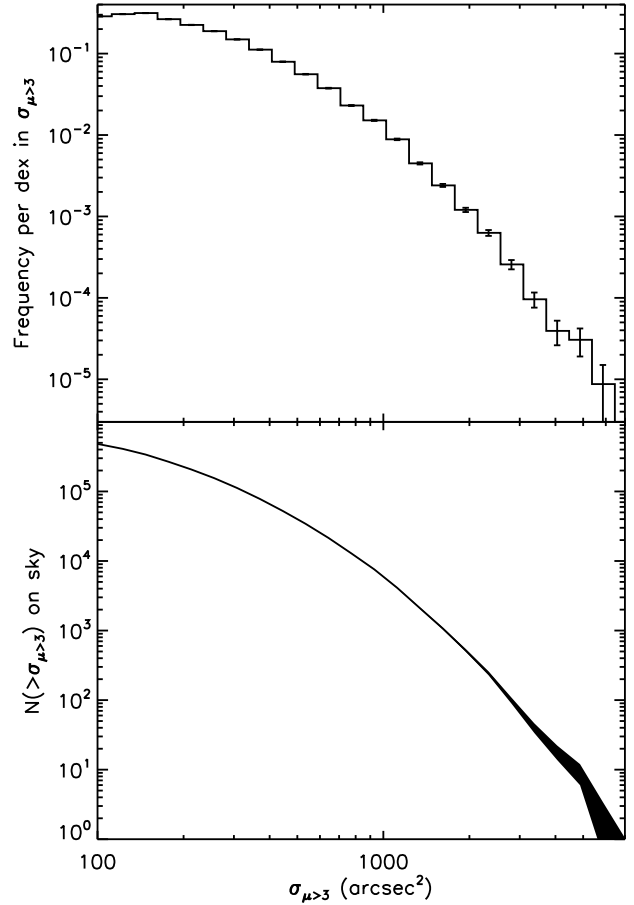}
\caption{Top: Frequency per dex in \sm of beams versus \smd. Bottom: Number of beams on the sky with \sm greater than the given value. Shaded regions represent 1$\sigma$ errorbars. We expect $477 \pm 21$ beams with \sm $>2000$ arcsec$^2$ over the whole sky. These beams should be among the best lensing beams based on theoretical modeling and observed lines of sight.}
\label{smfreq}
\end{figure}

The overall frequency per \sm can be seen in Figure \ref{smfreq}. The frequency is determined by calculating \sm for a selection of random beams from MXXL, as discussed in \ref{samples}. \sm bins in Figure \ref{smfreq} are spaced equally in logarithmic space. Frequency is plotted per dex in \sm to make this axis independent of bin size. 
The bottom panel of Figure \ref{smfreq} shows the number of beams on the sky with \sm greater than the given value. 
We expect to see $6025 \pm 78$ beams with \sm$>1000$ arcsec$^2$, and $477 \pm 21$ beams with \sm$>2000$ arcsec$^2$ on the sky. This error is due to both Poisson noise and cosmic variance in the simulation. The Poisson noise on the number of beams in each \sm is calculated to be the standard deviation on a Poisson distribution with that number as the mean. Cosmic variance is estimated by calculating the variance in results among origin points in the sample. 

By comparing the volumes available, one could expect to find $86 \pm 4$ beams with \sm$>2000$ arcsec$^2$ in the SDSS LRG catalog \citep{Ahn2012}. We expect $1102 \pm 14$ beams with \sm $>1000$ arcsec$^2$. The 200 LRG-dense beams found by \citet{Wong2013} in the SDSS thus are likely to be of the same class of beams, with typical \sm values over $1600$ arcsec$^2$.

We address the sensitivity of these results to the choice of cosmology in Section \ref{cosmology}.

\subsection{Best Lensing Configurations}

\subsubsection{Redshift Range}
\label{zcumuu}

The first beam configuration property we consider is the redshift range over which halos act as the most effective lenses for $z\sim10$ sources.
For practical observing reasons, we wish to find a limit in redshift where massive structures will no longer contribute significantly to lensing $z\sim10$ sources. \citet{Hennawi2007} calculate the number of giant arcs greater than 10'' for lenses out to $z$=1.5 for source planes of $z$=1-4. We perform a similar analysis, but for a \zten source plane, and use the change in \sm to indicate the importance of lenses in each redshift bin. 

There are three main effects that determine the redshift range where good lensing clusters lie. The first is the availability of massive halos, given their formation time. At higher redshifts, the most massive clusters will not have had time to assemble. The second is the effect of volume observed in a lightcone, which grows with redshift. The third is the redshift dependence of the strong lensing cross section, which is related to \smd. Depending on the model assumed for the halos, this will go as $D$ or $D^2$, where $D=\frac{D_{ol} D_{ls}}{D_{os}}$, the ratio of the angular diameter distances between the observer ($o$), lens ($l$), and source ($s$). This quantity peaks at a redshift of $z=0.5$, and drops slowly off from there. 
We wish to study not only the redshifts which are effective for finding single-halo lenses, but also for multi-halo beams.

In order to study the effect on \sm of only observing galaxies out to a certain redshift, we calculate \sm for each beam, at various maximum lens redshifts, using {\it lensmodel} as before.
We focus on  high \sm beams, so we do this analysis for all beams with a total \sm of $>1000$ arcsec$^2$.
 We include beams with any number of halos, so beams with two halos at different redshifts will contribute to $d\sigma_{\mu>3}/dz$ at two different redshifts.

\begin{figure}
\includegraphics[height=2.5in]{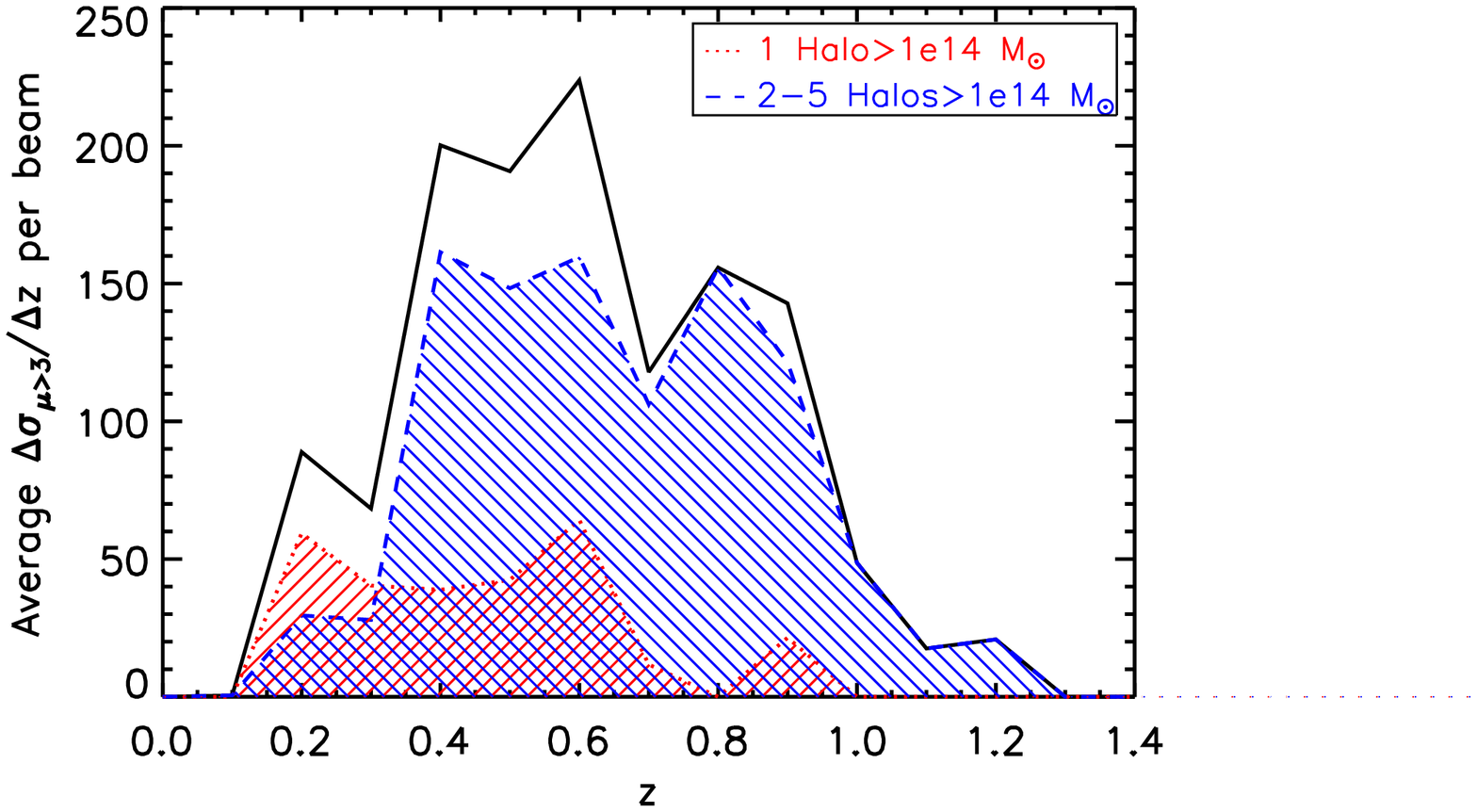}
\caption{Total change in \sm with redshift (black solid line). The y-axis shows the average change in \sm as the highest redshift considered is increased by 0.1, for each beam with \sm $>1000$ arcsec$^2$. The distribution along the x-axis shows the range of redshifts that contribute the most to \smd. The (red) dotted line shows the component of this curve from single-halo beams, and the (blue) dashed line shows the component for multi-halo beams. The single-halo beams with high \sm ($>1000$ arcsec$^2$) are typically found between redshifts of 0.1 and 0.6, but the good multi-halo beams are spread over a wider range of redshifts, 0.3-1.0.}
\label{zcumuxxl}
\end{figure}

The change in \sm for each redshift step is averaged for all beams with \sm $>1000$ arcsec$^2$ and plotted in Figure \ref{zcumuxxl}. 
This plot is then divided into two components, single-halo beams and multi-halo beams. The single-halo beams with high \sm typically lie between redshifts of 0.1 and 0.6. This is consistent with both theoretical results \citep{BartelmannMatthias1998}, and the observed locations of known clusters that produce giant arcs \citep{Hennawi2008}. However, the good multi-halo beams are spread over a wider range of redshifts, 0.3-1.0, which can be understood by looking back to one of the factors in the redshift dependence of \smd. For multi-halo beams, a high total mass can be obtained from two or more smaller (but still cluster-scale) halos, which means that high \sm beams are not constrained to epochs where the most massive clusters have formed. $10^{14} M_\sun h^{-1}$ halos are found over a wider redshift range than $10^{15} M_\sun h^{-1}$ halos. 

For beams with \sm$>1000$ arcsec$^2$, \sm increases by less than 5\% when the mass beyond $z=1.0$ is included, and by less than 1\% beyond $z=1.1$. Ignoring mass beyond these redshifts will result in an underestimation of \smd.

\subsubsection{Number of Halos in the Beam}
\label{nhalossec}
\begin{figure}
\includegraphics[height=2.5in]{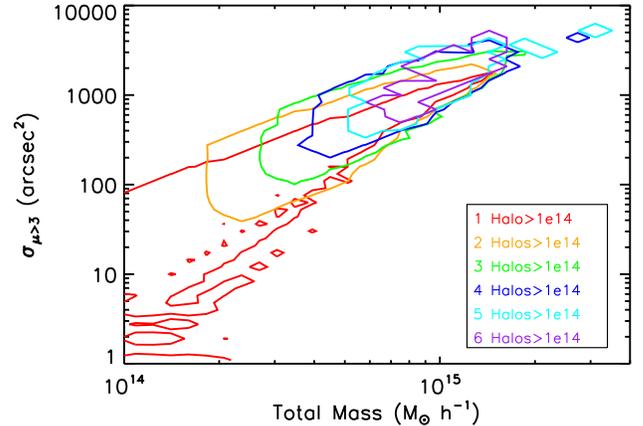}
\caption{Total mass in beam vs. \sm broken down by number of halos in the beam. Contours containing 99.7\% of the points are colored by the number of cluster-scale (mass greater than $10^{14} M_\sun h^{-1}$) halos in the beam. For a given mass, both the upper envelope and median \sm increase with the number of halos in the beam. This result is confirmed statistically by a multivariate analysis, which shows a significant residual correlation between \sm and the number of halos after controlling for the total mass in the beam.}
\label{ncontour}
\end{figure}

 In \citet{Wong2012}, beams whose same total mass is distributed into more projected halos tend to have higher \smd, i.e., are better lenses. We explore this idea with the MXXL in Figure \ref{ncontour}, in which the data from Figure 3 are broken down by the number of halos in each beam. At a given mass, \sm increases with the number of halos, for both the median and upper envelope of each distribution. Thus, our MXXL analysis suggests that distributing a beam's mass among more halos improves the lensing.

To ascertain the statistical significance of this result, we must control for the total mass: because the mass function is steep and we have a minimum halo mass in the sample, 5-halo beams tend to have more mass than 1-halo beams.  We know from Figure \ref{msigma} that there is a strong correlation between \sm and total mass, so we must fix mass to quantify the additional contribution to \sm when that mass is split among multiple halos. Therefore, we perform a multivariate analysis on \smd, total mass $M$, and number of halos $N_h$ with mass greater than $10^{14} M_\sun h^{-1}$. First, we calculate the Spearman rank correlation coefficient $r_{ij}$ for each pair of variables. The partial correlation coefficient $p$ for $N_h$ and \sm is then calculated as
\begin{equation}
p_{\sigma_\mu, N_h} = \frac{r_{\sigma_\mu,N_h} - r_{N_h, M} r_{\sigma_\mu, M}}{\sqrt{(1-r_{N_h, M}^2)(1-r_{\sigma_\mu, M}^2)}}.
\end{equation}

The quantity $z$, and its variance $\sigma_z^2$ are often used to describe this statistic, because $z$ will follow a normal distribution. These quantities are defined as,
\begin{equation}
z=\frac{1}{2} ln \frac{1+p}{1-p}; \ \ \ \sigma_z^2 = \frac{1}{N-1-k}
\end{equation}
where $N$ is the sample size, and $k$ is the number of variables considered. Here, $N=10^6$ and $k=3$. 

We use this analysis of the partial correlation coefficients to determine the residual correlation between \sm and $N_h$ in MXXL:
\begin{equation}
p_{\sigma_\mu,N_h} = 0.064 \ \ \ (z=64 \sigma_z),
\end{equation}
which indicates a correlation between \sm and the number of halos even at fixed mass. While much of the statistical power comes from beams with low \smd, the residual correlation remains highly significant ($z=8\sigma_z$) for the subsample of beams with \sm$>2000$ arcsec$^2$.

Following \citet{Wong2012}, we explain the residual correlation of \sm with number of halos (controlling for mass) in terms of interactions among the lensing potentials of the halos. Even when there is no physical interaction among halos, the projected lensing potentials can overlap in a way that boosts the magnification and thereby enlarges the region with $\mu > 3$.  We note that the residual correlation is seen only when we consider halos of mass $10^{14} M_\sun h^{-1}$ and above in the MXXL.  If we include halos of $10^{13} M_\sun h^{-1}$ and above, the residual relation changes to an anti-correlation:  separating the mass into more, but generally less massive, halos now produces lower \smd.  This change arises because our beam size is fixed and because less massive halos are smaller, worse lenses.  For a given beam size, massive halos can be close enough in projection for their lensing potentials to interact, but less massive halos may not be.  If our analysis is repeated with a smaller fixed beam size, the benefit of breaking the mass into smaller halos extends down to lower masses.

\begin{figure}
\includegraphics[height=2.5in]{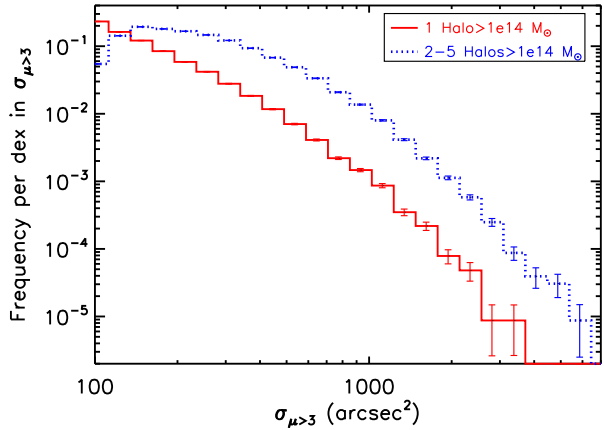}
\caption{Frequency of beams per dex in \sm vs. \smd, broken down into single-halo and multi-halo beams. The best lensing beams, at high \smd, are dominated by multi-halo beams. Beams with \sm$>2000$ are ten times more likely to contain multiple halos than single halos.}
\label{dnplot2}
\end{figure}

Additionally, we want to know if the highest \sm beams are likely to contain multiple halos. We break up the plot of frequency per \sm shown in Figure \ref{smfreq} by number of halos in each beam. This frequency analysis can be seen in Figure \ref{dnplot2}. For the highest \sm bins, multi-halo beams are more common than single-halo beams. This is due to the combination of the likelihood of compiling more lower-mass halos, the wide redshift range available for effective lensing, and the high frequency of line-of-sight mass, which boosts the lensing power of already massive halos (discussed further in Section \ref{clashcompare}). For beams with \sm$>2000$ arcsec$^2$, multi-halo beams are ten times more common than single-halo beams.
The frequencies of multi-halo beams at the high \sm end can be compared to that of the whole sample of non-empty beams (any \sm), which is 81\% single-halo beams, 16\% 2-halo beams, 2\% 3-halo beams, and less than 1\% 4-6 halo beams. Although beams with fewer halos are more common, beams with more halos are more likely to have high values of \smd.
These results suggest that surveys for the best gravitational lenses are underestimating the amount of line-of-sight structure of the best lensing clusters, and may be missing or mischaracterizing most of the best gravitational lenses on the sky.

\subsubsection{Separation of Halos in Redshift and on the Sky}

\begin{figure}
\includegraphics[height=2.5in]{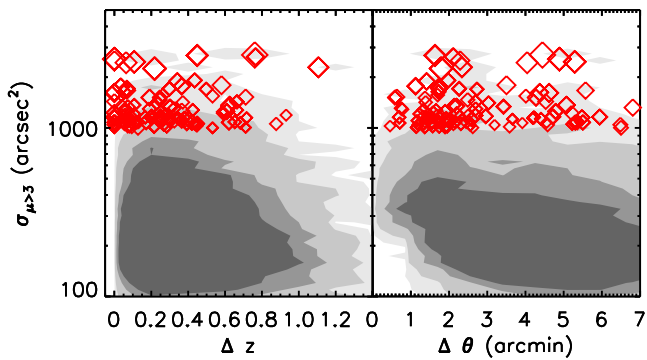}
\caption{Distribution of \sm for two-halo beams in redshift separation and projected angular separation. Beams with \sm$>1000$ arcsec$^2$ are plotted individually, and the total sample is shown as contours. Contours are at values of (1,10,50,100)*0.008 per redshift per $log$ arcsec$^2$ for $\Delta z$ and (1,10,50,100)*0.03 per arcminute per $log$ arcsec$^2$ for $\Delta \theta$. Point diameter is linearly proportional to the total mass of the beam.  High \sm beams are distributed roughly evenly with $\Delta z$, but clump around $\Delta\theta\sim2$ arcminutes. A multivariate analysis shows a significant anti-correlation of \sm with $\Delta\theta$, and no significant correlation of \sm with $\Delta z$.}
\label{dzdtheta}
\end{figure}

To study how other configuration quantities affect \smd, we consider a subset of the MXXL beams with only two $>10^{14} M_\sun h^{-1}$ halos. These beams are the simplest test cases for studying how the lensing potential interactions described in the previous section are affected by the halo separation in redshift ($\Delta z$) and on the sky ($\Delta\theta$). We plot contour-scatter plots of these quantities against \sm in Figure \ref{dzdtheta}. 
There are fewer beams as $\Delta z$ increases, because of the dropoff in massive halos at larger redshift. 
The projected separation of the halos was also considered in \citet{Wong2012}, who found a peak in \sm for $\Delta\theta$ values of $\sim100$ arcseconds. However, this peak varied with the total mass in the two halos, and because we do not fix the total mass as they did (at $2\times10^{15} M_\sun h^{-1}$), our results include the sum of many peaks for a range of total masses. The highest \sm beams do cluster at $\Delta\theta\sim2$ arcminutes, or 120 arcseconds. Some of the high \sm beams do not follow this trend, and are driven to high \sm by their high total masses. This can be seen in Figure \ref{dzdtheta}, where the effect in mass can be seen by the plot point sizes. Some beams with separations close to $\Delta\theta\sim2$ arcminutes have high values of \sm for lower total masses than high \sm beams at larger separations.

We study these trends quantitatively using the multivariate analysis technique used in the previous section. Controlling for total mass and projected angular separation, \sm has no significant correlation with $\Delta z$. However, controlling for total mass and redshift separation, \sm has a $3\sigma$ significant anti-correlation with $\Delta\theta$.  
These results reinforce the visual trends. For redshift separation, the high \sm beams are roughly distributed across the $\Delta z$ range considered. Many of these beams have a massive halo at low redshift, and a less massive halo at some other redshift. Since the lower \sm beams drop off with increasing $\Delta z$, this results in an overabundance at higher $\Delta z$. The anti-correlation observed in $\Delta\theta$ reflects the clump at $\sim2$ arcminutes.

We do not study the variation in \sm due to changing the concentration and ellipticity of the halos. We are ultimately concerned with finding how \sm varies with parameters that can be determined from observations in current large scale surveys, such as the SDSS \citep{Ahn2012}. Using luminous red galaxies (LRGs) as mass tracers, we can use the number of LRGs or their total luminosity as a proxy for total mass and the distribution of line of sight structure \citep{Wong2013}. Constraints on concentration and ellipticity will only come after mass modeling of the beam in followup observations and strong and weak lensing analyses.

\subsection{Observational Identification of the Best Beams}
\label{selection}

\begin{figure*}
\includegraphics[height=5in]{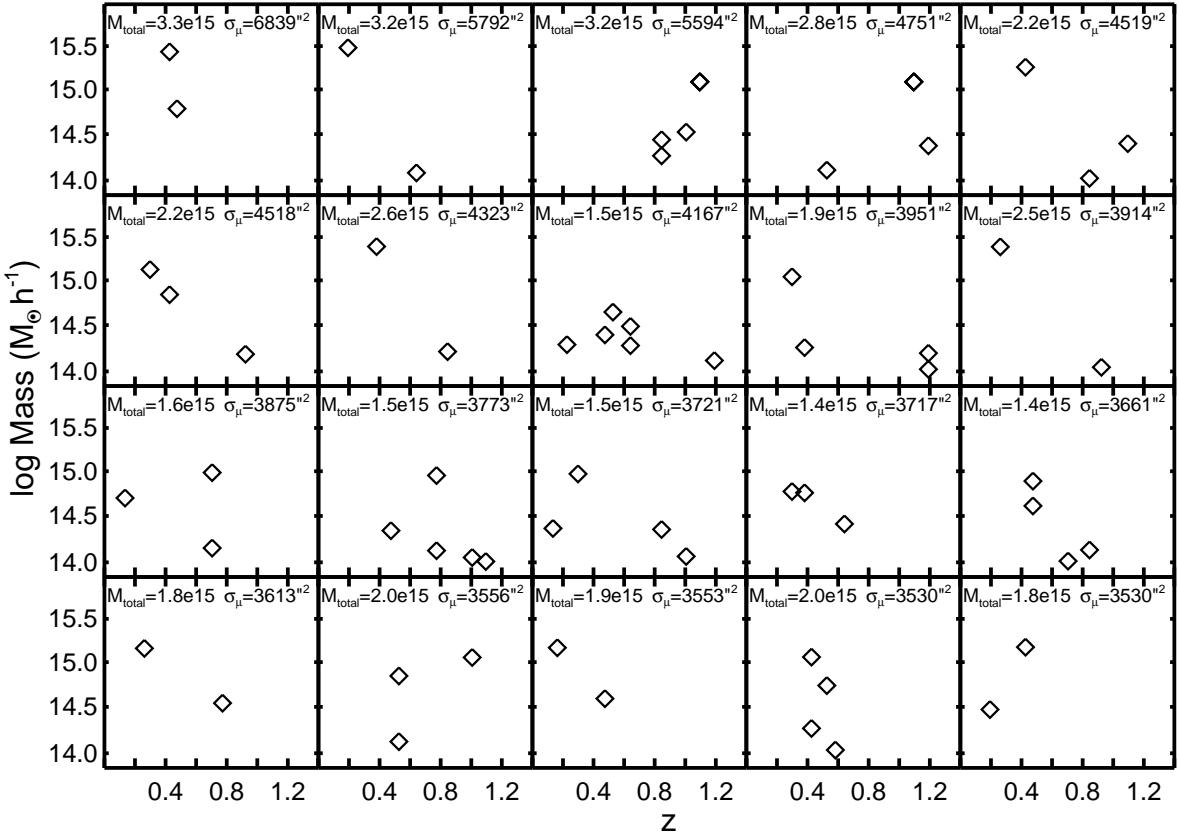}
\caption{Mass vs. redshift plots of top 20 \sm beams in MXXL sample. These are plots of the mass and redshift of halos within the top beams, with \sm (arcsec$^2$) and total mass ($M_\sun h^{-1}$) listed in each plot.}
\label{postagestamps}
\end{figure*}

We would like to know which observable beam properties result in a higher probability of a beam having a high \smd. Mass-redshift plots of the best beams in our MXXL sample are shown in Figure \ref{postagestamps}. We want to select these top \sm beams using only information we might obtain observationally from these beams, with a minimal amount of additional follow-up telescope time. 

We begin by selecting the quantities we would be likely to observe for a large sample of possible lensing beams. Total LRG luminosity can be used as a proxy for total mass. With photometric redshift information, we could estimate the redshifts of the component LRGs, and perhaps define cluster-scale halos. Though the scaling between number of LRGs and halo mass has significant scatter \citep{Ho2009,Zheng2009b}, this method allows a wide section of the sky to be searched. Other methods for finding the number of halos in a beam include comparing the LRG luminosity proxy for total mass to the Sunyaev-Zeldovich \citep{Sunyaev1980} measurement, as the S-Z effect measured will scale faster than linearly with mass, or using red-sequence fitting to further divide up the observed beam components. We adopt the following five quantities as potential observables: total mass, mass of the dominant halo in the beam, number of halos, minimum redshift of halo components, and maximum redshift of halo components.

We use a classification tree method similar to \citet{Richards2011} to determine a sequence of observational cuts that would generate the purest sample of high \sm beams. We define purity as the number of high \sm ($>2000$ arcsec$^2$) beams divided by the total number of beams selected into a bin. For the five observables outlined above, we choose a series of possible cuts, with 6-12 possible cuts per observable. The coarseness of the cuts is acceptable due to the large errors that would be present in their measurement. For each node in the tree, we perform the following procedure. We choose an observable parameter, and an accompanying parameter cut at random, and measure the purity of \sm$>2000$ beams for either side of the cut in parameter space. We do this 100 times for each node, and choose the parameter cut that results in the highest purity region of parameter space. Here, we grow the tree two levels. A parameter is not eliminated, even if it has just been chosen in the node above. We use 100 randomly selected training sets, and select the most commonly produced tree.

\begin{figure}
\includegraphics[height=3.5in]{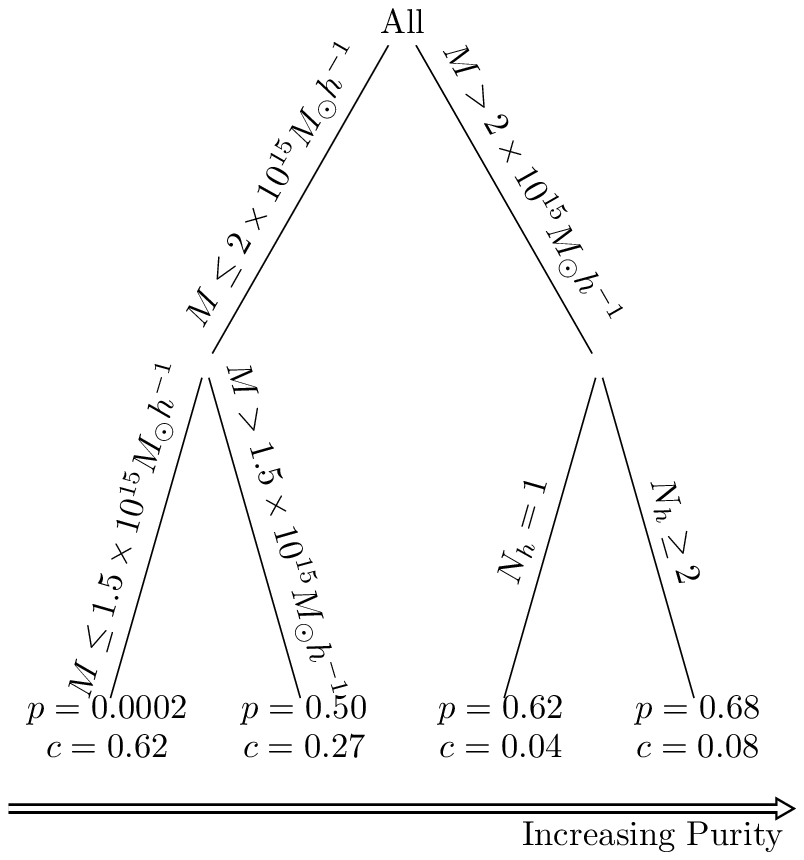}
\caption{Classification tree to find regions in parameter space of observables (total mass $M$, number of halos $N_h$, halo minimum and maximum redshifts, and mass of primary halo in beam) that maximize purity of the best \sm ($>2000$ arcsec$^2$) beams. Purity $p$ and completeness $c$ of each resulting node are listed. To create the most pure sample of high \sm beams, it is most helpful to select beams with total mass above $2\times10^{15} M_\sun h^{-1}$, made up of at least 2 structures along the line of sight. Cuts on all parameters are tested for each node. Selecting on the mass of the most massive halo or redshift cuts is not preferred, and so do not appear in the tree. For our sample, the highest purity bin produces 39 high \sm beams. Completeness adds to 1.01 due to rounding.}
\label{tree}
\end{figure}

A picture of the resultant tree is in Figure \ref{tree}. We learn several things from this method. The first is that, as expected, total mass is the most important parameter for maximizing purity. Having a high number of halos can additionally select a more pure sample. The highest purity region of parameter space for good beams is when the total mass is greater than $2\times10^{15}M_\sun h^{-1}$, and there are three or more halos in the beam. This selection method must be used carefully to avoid selecting only halos with mass overestimates due to systematic errors. For our sample, the highest purity bin produces 39 high \sm beams.

We can also learn from the features that are not selected in the tree. The mass of the dominant halo is not as useful as total mass or number of halos. This is important, as this selection is essentially what is currently employed to find good gravitational lenses. 

We have focused so far on sample purity, as we want to optimize the telescope time available for studying good beams. However, cuts to the sample necessarily sacrifice completeness, and good beams may be missed. In Figure \ref{tree}, we list both the purity $p$ and completeness $c$ in the selection nodes. 

\subsection{Likely LOS Mass for Traditional Strong Lenses}

\label{clashcompare}

We have established that some of the best lensing beams in the universe will have multiple massive halos. However, current known strong lenses are generally selected to be a single massive halo, and any mass along the line of sight (LOS) is often neglected in the lensing analysis and mass reconstruction. We investigate  how these beams may be affected by typical amounts of LOS mass. We draw four clusters from the Abell and CLASH \citep{Postman2012} cluster samples, and find analogue halos in MXXL with similar masses and redshifts ( Table \ref{clashtable}). We select clusters at the closest redshift snapshot to the real clusters, and with masses spanning the range of the stated uncertainties in the masses of the real clusters.

\begin{figure}
\includegraphics[height=2.5in]{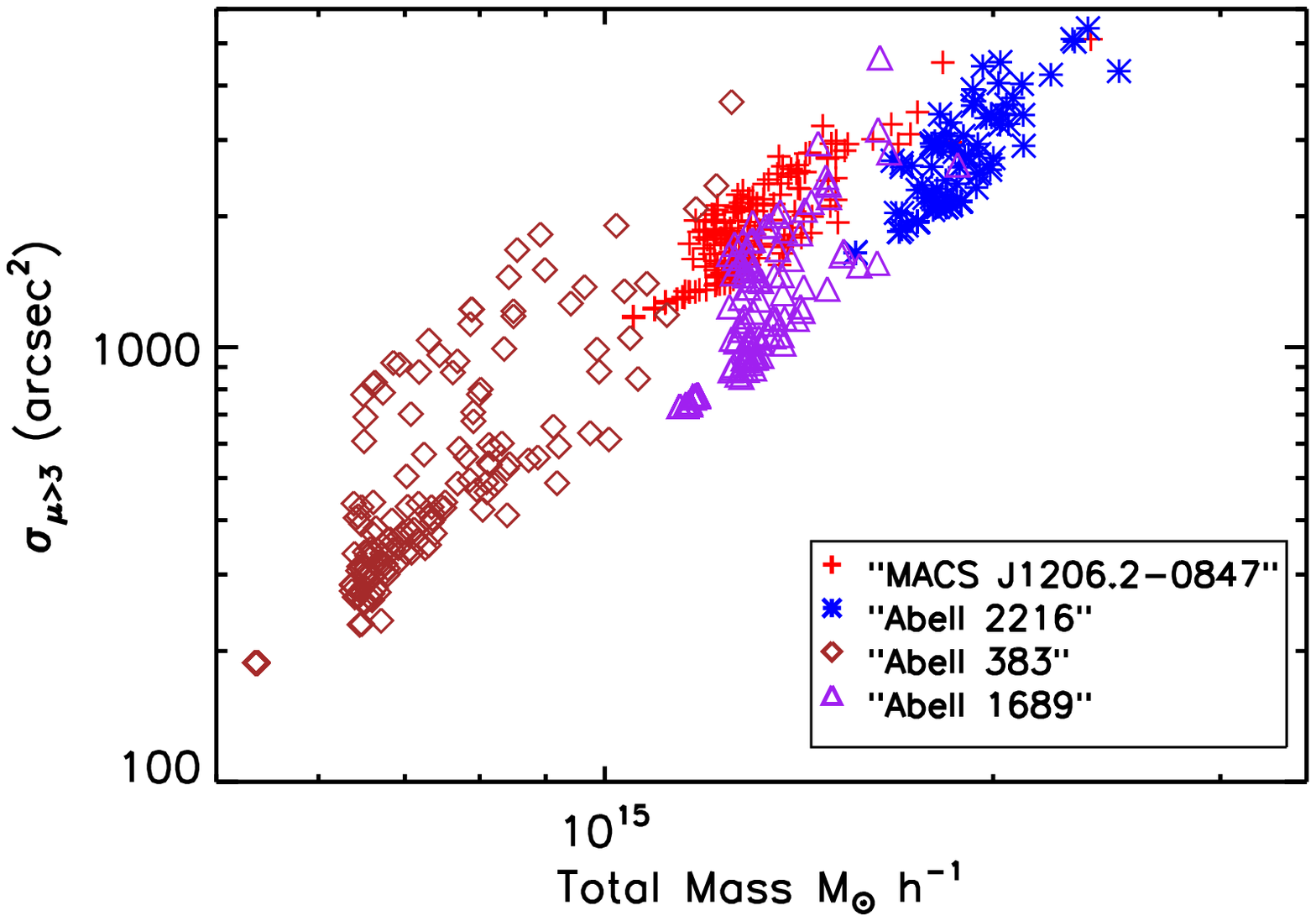}
\includegraphics[height=2.5in]{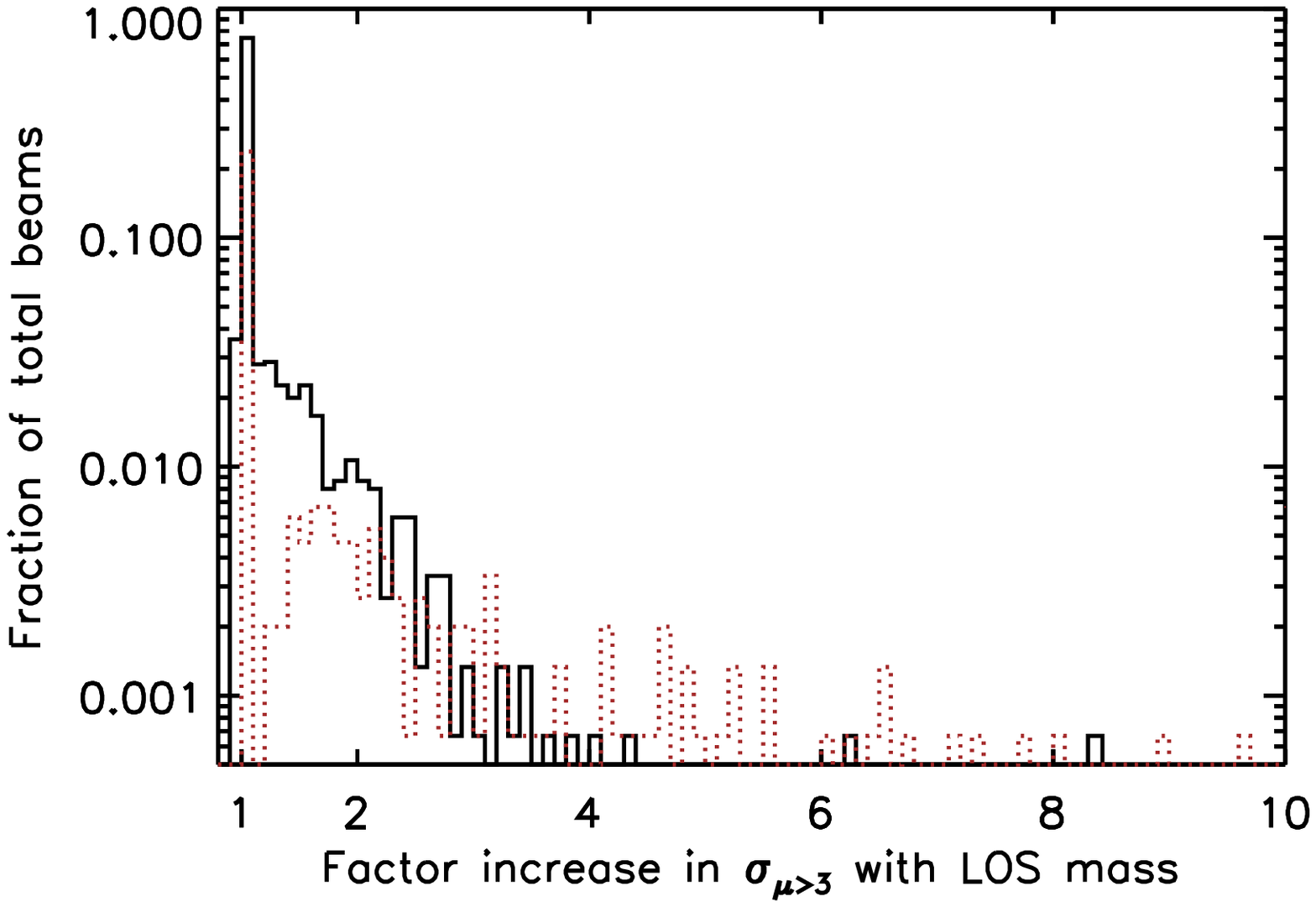}
\caption{Comparison of CLASH cluster analogues alone vs. including likely cluster-scale LOS mass. Top: \sm vs. total mass.  Different symbols indicate different main halos. Diamond points are analogues of Abell 383, plus points of MACS J1206.2-0847, triangles of Abell 1689, and asterisks of Abell 2216. Values for the cluster analogue with no LOS mass are those in the bottom left of each group. Bottom: Histogram of the multiplicative effect of including LOS mass on \smd. 1 represents no change. Beams include at least one additional cluster-scale ($M_{200}>10^{14}M_\sun h^{-1}$) halo 28\% of the time. In 20\% of the total cases, \sm increases by more than 25\%, and in 10\% of the total cases, \sm increases by more than 80\%. In this plot, we show the histogram for the three most massive comparison clusters in black. Due to its low mass, Abell 383 is disproportionately affected by LOS structure, and is overplotted as a thinner orange line.}
\label{clash}
\end{figure}

For each simulation cluster, we create a sphere of possible origin points around the halo the appropriate redshift distance away. Beams are chosen from this sphere of origin points, forced to go through the halo in consideration, and any other halos that fall within this beam are considered in the calculation of \smd. We perform this analysis for five analogue halos for each of the four comparison clusters.
How much and how often \sm is affected by the LOS mass can be seen in Figure \ref{clash}. 
Additional LOS mass causes a systematic increase in \smd, which serves as a proxy for how the magnification map would change.
Beams include at least one additional cluster-scale ($M_{200}>10^{14}M_\sun h^{-1}$) halo 28\% of the time. In 20\% of the total cases, \sm increases by more than 25\%, and in 10\% of the total cases, \sm increases by more than 80\%.
Even when excluding LOS mass halos greater than the original lens, \sm boosts can still be larger than 200\%.

These results are consistent with previous studies of LOS mass, which have found a general increase in measures of lensing power with LOS mass. \citet{Puchwein2009} find that lensing cross-sections (defined differently than \smd) can increase by up to 50\% due to additional structure along the LOS, with this effect increasing with source redshift. We observe stronger boosts in \sm due to the higher source redshift of $z\sim10$, and the fact that our definition of \sm is more sensitive to regions of intermediate magnification. Our results are a conservative estimate of the total number of beams affected by LOS mass, as we do not consider halos below our previous mass cut of $10^{14} M_\sun h^{-1}$. If lower mass halos are included, they will increase the total number of beams affected, but we would not expect a large contribution to the fraction of dramatic \sm boosts.

Due to the X-ray selection technique used to select the CLASH \citep{Postman2012} clusters, the real CLASH clusters are less likely to have LOS mass than those selected via other methods. X-ray selection techniques are less likely to be influenced by less massive LOS mass than optical selection techniques, so the MACS clusters are likely to only contain a single massive halo \citep{Ebeling2010}. Additionally, the real Abell 1689 has a higher concentration than our mass-redshift relation would predict. As the analysis in \citet{Wong2012} shows, higher concentrations can increase \sm for massive halos. The effect of concentration scatter on our results is studied in Section \ref{haloparams}.

In cluster-scale strong lensing studies, LOS mass has been found to shift the image positions by one to several arcseconds. \citet{Jullo2010} perform a similar analysis, obtaining lines of sight using the MS, and find that in most cases the LOS mass shifts the lensed image position, adding about 1 arcsecond to the total error budget of the lens model. \citet{Host2012} finds that LOS mass shifts the positions of lensed images by several arcseconds, with the error increasing with source redshift. They find these errors to be highly correlated and the dominant source of systematic errors in lens modeling. Even if a mass model could be constructed that accurately reproduced the lensed image positions, the mass profile of the cluster measured would include these systematic errors. 

Because of the potential for large changes, and the systematic nature of this effect, we argue that LOS mass should be considered in the modeling of these lenses, and that it can be used to the advantage of magnifying high redshift galaxies. 

\begin{table*}
\caption{Known lensing cluster comparisons}
\label{clashtable}
\begin{tabular}{|l l l l l l|}
\hline
\hline
Cluster & $M_{vir}$ (obs) $(10^{15} M_\sun h^{-1})$ & Redshift (obs) & Reference & $M_{vir}$ (sim) $(10^{15} M_\sun h^{-1})$ & Redshift (sim) \\ 
\hline
Abell 1689 & $1.3\pm0.4$ & 0.1832 & \citet{Lemze2009} & 1.14-1.18 & 0.17\\ 
MACS J1206.2-0847 & $1.1\pm0.2$ & 0.439 & \citet{Umetsu2012} & 1.05-1.15 & 0.46 \\ 
Abell 2216 & $1.7\pm0.2$ & 0.225 & \citet{Coe2012} & 1.5-1.9 & 0.24 \\ 
Abell 383 & $0.537^{+0.070}_{-0.063}$ & 0.189 & \cite{Zitrin2011a} & 0.536-0.538 & 0.17 \\
\hline
\end{tabular}
\end{table*}

\subsection{Detecting \zten Sources}
\label{ndetectsec}

The ultimate measure of a lensing beam is the number of detections per magnitude. To determine the number of detections, we must assume a luminosity function at $z\sim10$,  a distribution of source sizes and shapes, and an example observing program. Because of the large uncertainties in these quantities, we have employed \sm as our metric so far.

Nevertheless, here, we assume a Schechter function, parameterized by the faint-end slope $\alpha$, a characteristic magnitude $M_{UV}^*$, and a normalization factor $\phi^*$. We take $\alpha=-1.73$, $M_{UV}^*=-17.7$, and $\phi^*=1.15\times10^{-3}$, from the constraints in \citet{Oesch2013}.

The distribution of source angular sizes is determined by taking the distribution from $z\sim7$ sources, and evolving with redshift. This size evolves as $(1+z)^{-1.28}$ \citep{Ono2012}. We assume a normal distribution for the sizes. The sizes convert to angular sizes of $0.11''\pm0.4''$. The shape of the sources is also varied, with major-to-minor axis ratios of $1.2 \pm 0.3$, normally distributed.

We assume a sample observing program using a 7'$\times$7' mosaic (about 8 times the area of the \citet{Bouwens2011a} blank field). We use the PSF of HST/WFC3, 0.15'', and assume that we can reach the same magnitude limits as \citet{Bouwens2011a}, 29.8 AB mag in the $H_{160W}$ filter. There is a tradeoff here between a wide survey and the depth attained. For constraining the faint end of the luminosity function, very deep observations are preferred. 

\begin{figure}
\includegraphics[height=2.5in]{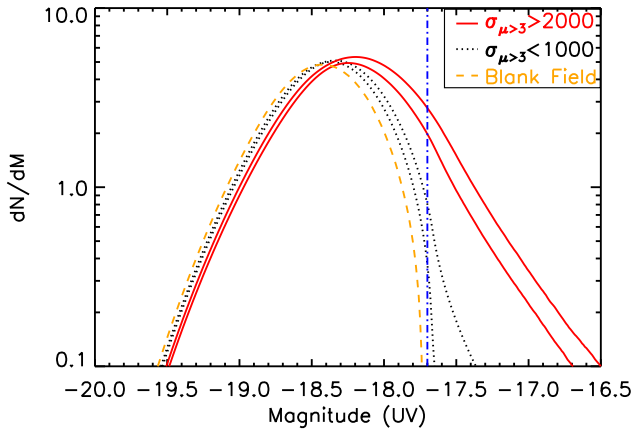}
\caption{Number of detections gained per magnitude, $dN/dM$, vs. magnitude $M_{UV}$. The region outlined by solid (red) lines represents the beams with \sm $>2000$ arcsec$^2$, and the region outlined by black dotted lines represents the beams with \sm $<1000$ arcsec$^2$. Scatter within these bands is due to various magnification distributions within each beam, and varying values of \smd. For reference, $M^*_{UV}$ is plotted as a dash-dotted (blue) line. The higher \sm beams result in more detections, and can push down towards fainter detections. This calculation requires assuming a luminosity function at \zten \citep{Oesch2013} and a hypothetical observing program (details in text). We also compare our calculations to an equivalent blank field (orange dashed line) observing program of the same FOV and limiting magnitude. Beams with \sm $>2000$ arcsec$^2$ will push further down the luminosity function than blank fields.}
\label{dndm}
\end{figure}

Using this method, we can calculate the number of \zten galaxies that would be detected using each lensing beam pulled from the simulations. Both the number of detections and number of detections fainter than $M_{UV}^*=-17.7$ correlate with \smd. 
We calculate the gain in number of detections with magnitude $dN/dM$ for each beam. Averages of $dN/dM$ for two ranges in \sm are plotted in Figure \ref{dndm}. Beams with high \sm allow the detection of more sources, and can push further down in magnitude. We also compare our calculations to a hypothetical blank field observing program of the same size and limiting magnitude.  
Beams with \sm$>2000$ arcsec$^2$ find 1.1-1.4 times more \zten detections than a blank field, and 100-250 times more detections fainter than $M_{UV}^*=-17.7$.

%%%%%%%%%%%%%%%%%%%%%%%%%%%%%%%%%%%%%%%%%%%%%%%%%%%%%%%%%%%%%%%%%%%%%%%%%%%%
%%%%%%%%% Discussion %%%%%% %%%%%%%%%%%%%%%%%%%%%%%%%%%%%%%%%%%%%%%%%%%%%%%
%%%%%%%%%%%%%%%%%%%%%%%%%%%%%%%%%%%%%%%%%%%%%%%%%%%%%%%%%%%%%%%%%%%%%%%%%%%%

\section{Discussion}
\subsection{Choice of Cosmology}
\label{cosmology}
One complication arises in using the Millennium simulations, which were conducted using a WMAP 1 cosmology. The largest discrepancy between the Millennium simulations and the most recent values is the clustering parameter $\sigma_8$. The Millennium simulations use a value of $\sigma_8=0.9$. The most recent value is $\sigma_8=0.829$ \citep{PlanckCollaboration2013}. While the Planck value of $\Omega_m=0.315$ is higher than the value $\Omega_m=0.25$ used in the simulations, the mass function predicted with the Millennium cosmology still overpredicts the number of massive halos at $z\sim0.5$. The number density is almost an order of magnitude higher in WMAP1 than Planck for halos with $M\sim10^{15.5} M_\sun h^{-1}$, the range of the most massive halos in our sample. 

To study how our results change for different cosmologies, we use the rescaling technique by \citet{Angulo2010}. The redshifts assigned to each simulation snapshot are changed to reflect the new cosmology. Lengths and masses are also rescaled, and we then create a new halo catalog \citep{Ruiz2011}.
Effectively, this method takes halos that would have been at higher redshift, and assigns lower redshifts to compensate for the longer assembly time due to the smaller value of $\sigma_8$. For the WMAP 1 to Planck cosmology scaling, a halo at redshift 0.68 will now be at $z=0.62$, halos at redshift 0.32 will shift to $z=0.26$, and those at redshift 0.12 will be at $z=0.05$. Any halos that have not formed by redshift 0.08 are dropped from the sample. Masses are scaled by a factor of 1.00548, and the length of the box is scaled by 0.93. 

\begin{figure}
\includegraphics[height=2.5in]{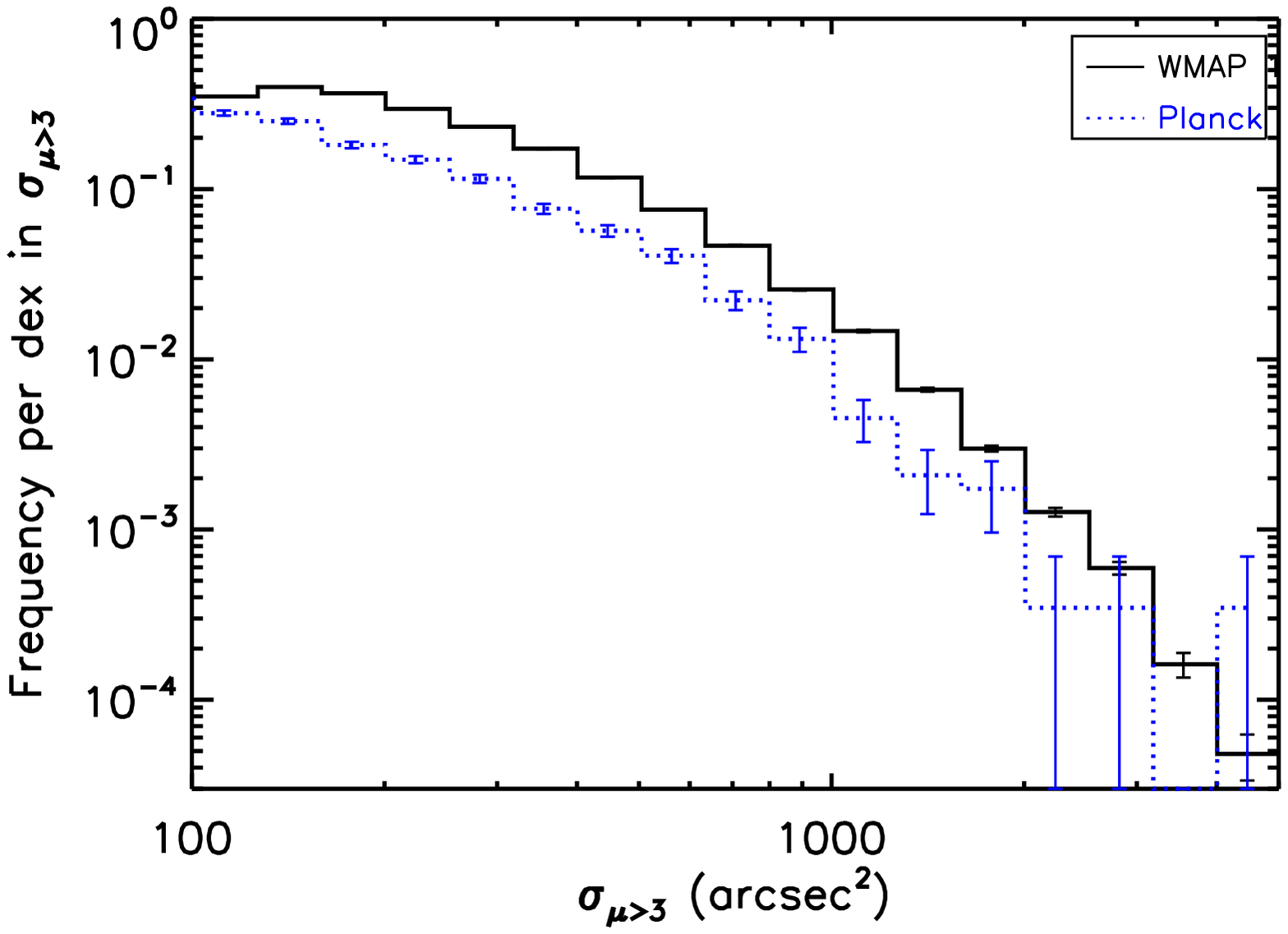}
\includegraphics[height=2.5in]{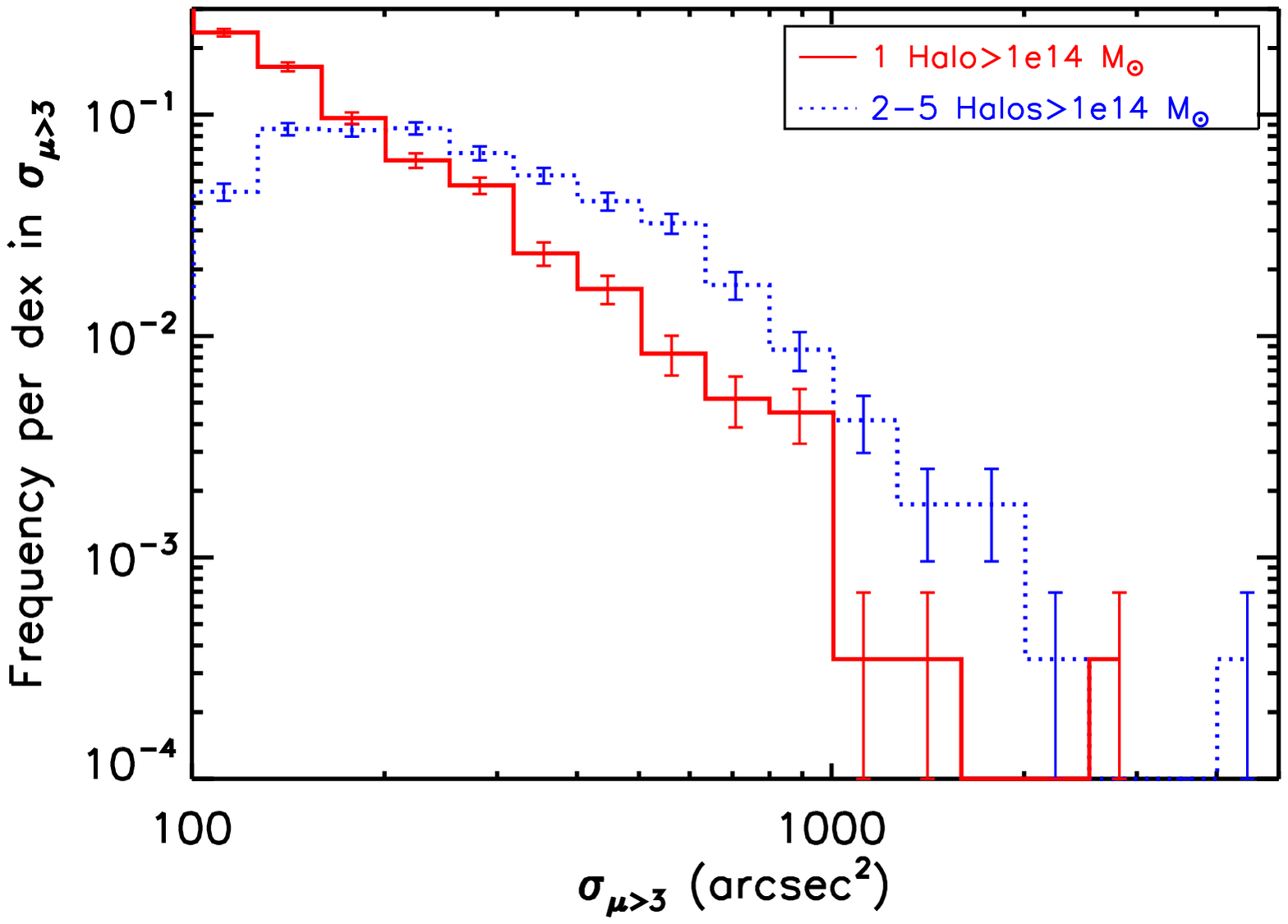}
\caption{Top: Frequency of beams per dex in \sm vs. \sm for different cosmologies. We rescale the MXXL simulation using the method in \citet{Angulo2010} to the results from the \citet{PlanckCollaboration2013}, which have a lower $\sigma_8$ and higher $\Omega_m$ than the cosmology used in MXXL. There are fewer massive halos at each redshift in the updated cosmology, which results in fewer high \sm beams. Bottom: Frequency of beams by \sm and number of halos with mass greater than $10^{14} M_\sun h^{-1}$ for the updated cosmology. Our previous result, that the best \sm beams are most likely to consist of multiple massive halos, still holds.}
\label{scaling}
\end{figure}

We calculate \sm for a sample of beams from the updated cosmology. We select 36000 random beams, a third of which contain at least one massive ($M>10^{14} M_\sun h^{-1}$) halo. The area probed by this sample is 490 sq degrees. Results can be seen in Figure \ref{scaling}. Beams from the updated cosmology have lower values of \smd. The number of beams with \sm$>2000$ arcsec$^2$ is reduced by 25\%. We examine whether our other conclusions are affected by the change in cosmologies. The bottom plot in Figure \ref{scaling} shows the updated version of Figure \ref{dnplot2}, where the frequency of beams by \sm is broken down by the number of halos. Despite the change in cosmology, the best \sm beams are still most likely to contain multiple massive halos. The redshift dependence of \sm still peaks between 0.2 and 1. As 33\% of random lines of sight still contain massive halos, lines of sight to known strong lenses still have a substantial chance of containing additional massive structures, but the likely fractional changes in \sm will decrease. The mean \sm in the scaled simulation is 18\% lower than the original. The quantitative changes to our results do not affect our overall qualitative conclusions.

\subsection{Choice of Beam Size}
Given the choice of $7^\prime\times7^\prime$ beam size, one might wonder if the significance of multi-halo beams is overestimated. We perform an analysis on a subset of beams, with a beam window of 3.5$^\prime \times$3.5$^\prime$. This is closer to the size of the HST ACS camera field of view. The probability of finding a massive halo in a beam goes down as the size of the beam decreases. As expected, the fraction of beams with at least one $10^{14} M_\sun h^{-1}$ halo is only 9\%, lower than the 30\% for the larger beams. 96\% of these beams have only one halo, and 4\% have multiple cluster-scale halos. Despite the reduced frequency, the highest \sm beams are still those with multiple halos. This indicates that the best lensing beams, even for smaller fields of view, will consist of multiple massive halos along the line of sight.

\subsection{Choice of Threshold Magnification}
Throughout, we have chosen a threshold magnification $\mu_t=3$. This value is based on the specific science goal of detecting \zten galaxies, and other values may be more appropriate for different science goals. 
We study the dependence of our results on $\mu_t$ by using $\mu_t=10$ to represent a version of \sm for high magnification regions. $\sigma_{\mu>10}$ is a similar metric to the arc-producing cross-section used in other studies. High magnification may be desirable if the science goals are to observe the faintest possible galaxies, lower redshift galaxies, or to find giant arcs. In this analysis, we continue to observe a residual correlation of $\sigma_{\mu>10}$ with number of halos, controlling for mass. The highest $\sigma_{\mu>10}$ beams are still primarily multi-halo beams. As $\mu_t$ is lowered, the lensing interaction between the halo potentials becomes more important, and multi-halo beams will continue to dominate.

For the specific science case of detecting faint \zten galaxies, the number detected (in the hypothetical observing program in the previous section) should correlate well with $\sigma_{\mu>\mu_t}$ for our choice of $\mu_t$. We determine the correlation between $\sigma_{\mu>\mu_t}$ and the number of detections using the Spearman rank correlation coefficient. This correlation is higher for lower values of $\mu_t$, and significant for all values of $\mu_t$ within 2-20, though relatively flat. This result, together with those from using $\mu_t=10$, demonstrate that our results are not sensitive to the choice of magnification threshold for various intermediate magnifications.

\subsection{Halo Shape Assumptions}
\label{haloparams}

We make several assumptions about the mass density profile, concentration parameter, and ellipticity of the halos.

Although detailed information about particle positions is intrinsically included in the simulations, it is not recorded for the halo catalogs. We therefore model each halo with an NFW profile \citep{Navarro1996}. While there will be variations from NFW profiles among the actual simulated halos, the NFW configuration is generally consistent with observations \citep[e.g.,][]{ Kneib2003,Shu2008} and used in recent lens modeling analyses \citep[e.g.,][]{Zitrin2013}. Thus, it remains our best choice for studying the likely lensing properties. 

In this paper, we use the concentration dependence on mass and redshift from \citet{Zhao2009}, with no scatter. We test the effect of scatter on our results by introducing a 0.14 dex scatter in the concentrations used for the halos in Section \ref{clashcompare}. For beams with \sm$>2000$ before or after the scatter is added, the mean change in \sm is consistent with no change, for both single-halo and multi-halo beams. Changes in \sm due to the scatter in concentration can range between 85\% and 130\% of the original value. While \citet{Wong2012} find that halos with high concentrations for their mass will have higher values of \smd, scatter in the relation will not bias our final results or observational selection criteria. If halo concentrations are known for a large catalog of massive halos, through weak lensing or other methods, it will be helpful to additionally select on halo concentrations to find the best lensing beams.

In this paper, we assume all halos are spherical, with no scatter in the ellipticity. From \citet{Wong2012}, we know that projected ellipticity of halos does not affect \sm as much as the concentration or other parameters, but that \sm can be boosted if the major axis is aligned along the line of sight to a halo. This effect is due to the change in projected concentration. \citet{Wong2012} find that the scatter introduced in the concentration from the ellipticity scatter is small, $\sim0.03$ dex, compared to the 0.14 dex scatter in the concentration relation. 

\subsection{Observations of Multi-Halo Beams}
Beams with multiple cluster-scale structures have been observed in the real universe. Spectroscopic observations \citep{S.MarkAmmonsKennethC.WongCharlesR.Keeton2013} of lines of sight selected to have high integrated LRG luminosities show two beams with multiple structures adding up to high total mass, and values of \sm between 2000-10000. \citet{Wong2013} identify 200 fields in the SDSS that are likely high \sm beams. Redshift histograms of these beams show more variation in the number of LRG redshift clumps than those of traditional lensing beams. These high LRG number beams have up to twice the number of LRGs per beam as traditional lensing clusters. These observational results are consistent with what we see here, that the best lensing beams are made of multiple massive structures along the line of sight.

%%%%%%%%%%%%%%%%%%%%%%%%%%%%%%%%%%%%%%%%%%%%%%%%%%%%%%%%%%%%%%%%%%%%%%%%%%%%
%%%%%%%%% Conclusions %%%%%% %%%%%%%%%%%%%%%%%%%%%%%%%%%%%%%%%%%%%%%%%%%%%%%
%%%%%%%%%%%%%%%%%%%%%%%%%%%%%%%%%%%%%%%%%%%%%%%%%%%%%%%%%%%%%%%%%%%%%%%%%%%%

\section{Conclusions}
We draw massive halos from sample lines of sight in the Millennium I \citep{Springel2005} and Millennium XXL \citep{Angulo2012} simulations. We define the quantity \smd, or \'etendue, the area in the \zten source plane where the magnification of the brightest image is greater than a threshold magnification of $\mu=3$. 
We characterize those beams with the highest lensing cross-sections \sm for detecting very high redshift ($z\sim10$) galaxies. 

Our main conclusions are:

\begin{itemize}
\item We calculate the frequency of high-\sm beams on the sky, in order to test whether these beams should exist in the real universe. The number of $7^\prime\times7^\prime$ beams with $\sigma_{\mu>3} >1000$ arcsec$^2$ out to a redshift of $z=1.4$, and including only halos with $M_{200}>10^{14} M_\sun h^{-1}$, is $6025 \pm 78$ over the whole sky. The number of beams with $\sigma_{\mu>3} >2000$ arcsec$^2$ is $477 \pm 21$. These numbers decrease by about 25\% if MXXL is scaled to a Planck cosmology.

\item We determine the characteristics of the high \sm beams. \sm  increases with total beam mass, and this relation can be fit by a power law with index of $1.36\pm0.02$.

\item The highest \sm beams with $\sigma_{\mu>3} >2000$ arcsec$^2$ are ten times more likely to have multiple cluster-scale halos than a single halo.
Beams with more halos have on average more total mass, and a resulting higher \smd. After controlling for total mass, we calculate a significant residual correlation between \sm and the number of massive halos ($>10^{14} M_\sun h^{-1}$), and an anti-correlation with the number of lower mass halos ($10^{13} M_\sun h^{-1}$). In other words, breaking up beam mass into several halos is beneficial to \sm provided the resulting halos have high cluster-scale masses. This is due to the tradeoff between decreasing lensing power for lower mass halos, and the likelihood of finding two or more close to each other within the specified beam size. These trends hold for different beam sizes, threshold magnifications, and cosmologies.

\item We study the effect of halo configuration for the special case of two-halo beams. Controlling for mass and redshift separation, we observe a $3\sigma$ significant anti-correlation of \sm with the angular separation between the halos. Good \sm  ($>1000$ arcsec$^2$) beam halos often have halo separations near 2 arcminutes. We observe no significant correlation between \sm and the redshift offset between the two halos.

\item Having determined characteristics of the best beams, we identify the observables that are most efficient at  {\it finding} the best beams. The redshift range that must be covered to find 95 \% of good \sm fields ($\sigma_{\mu>3} >1000$ arcsec$^2$) is $0<z<1.0$. We construct a classification tree to find regions in observable parameter space that have the highest purity of high \sm beams. We consider total mass, number of halos, mass of the most massive halo, and minimum and maximum redshifts.  It is most helpful to select beams with total mass above $2\times10^{15} M_\sun h^{-1}$, made up of at least 2 structures along the line of sight. Selecting on the mass of the most massive halo or on redshift is not as effective.

\item We put our analysis of high \sm beams into the context of other lensing surveys by comparing to known strong lens Abell 1689 and three CLASH \citep{Postman2012} clusters. For each lens, we select 5 analogue halos in mass and redshift from MXXL, and study possible lines of sight that include that halo. Additional line of sight (LOS) mass causes a systematic increase in \smd. Beams include at least one additional cluster-scale ($M_{200}>10^{14}M_\sun h^{-1}$) halo 28\% of the time. In 20\% of the total cases, \sm increases by more than 25\%, and in 10\% of the total cases, \sm increases by more than 80\%. Excluding LOS mass halos greater than the original lens, \sm boosts can still be larger than 200\%. Because of the potential for large changes, and the systematic nature of this effect, LOS mass must be considered in the modeling of these lenses.

\item The ultimate measure of beam quality is the expected number of \zten lensed galaxies one would expect to detect. This requires many additional assumptions over \sm. However, both the number of detections, and the number of detections fainter than $M_{UV}^*=-17.7$, are correlated with \smd. Beams with \sm$>2000$ arcsec$^2$ produce 1.1-1.4 times more \zten detections overall, and 100-250 times more detections fainter than $M_{UV}^*=-17.7$, when compared to a blank field.

\end{itemize}

Our results will inform searches for high \sm beams in the real universe, including with LSST \citep{Ivezic2008}. Our analysis of LRG number density and total LRG luminosity as tracers of massive halos has already shown such beams exist \citep{Wong2013}. Spectroscopic observations confirm them \citep{S.MarkAmmonsKennethC.WongCharlesR.Keeton2013}. We have also determined that even single-cluster lenses will be affected by line of sight structure. We argue that these same ``nuisance" systematics and uncertainties can be used to our advantage in detecting the earliest galaxies.

\acknowledgments
We thank the referee for their helpful comments, which improved this work. We would like to acknowledge the Millennium XXL team, and thank them for access to the XXL simulations. KDF thanks Dan Marrone, Dennis Zaritsky, Brenda Frye, Romeel Dav\'e and Brant Robertson for helpful discussions. KDF is supported by NSF grant DGE-1143953. KCW is supported by an EACOA Fellowship awarded by the East Asia Core Observatories Association, which consists of the Academia Sinica Institute of Astronomy and Astrophysics, the National Astronomical Observatory of Japan, the National Astronomical Observatory of China, and the Korea Astronomy and Space Science Institute. AIZ acknowledges support from NSF grant AST-0908280, as well as NASA grants ADP-NNX10AD476 and ADP-NNX10AE88G. AIZ also thanks the John Simon Guggenheim Memorial Foundation
and the Center for Cosmology and Particle Physics at NYU for their support. CRK is supported by NSF grant AST-1211385.

%\clearpage
\bibliographystyle{apj}
\bibliography{uberbeamsrefs8}

\begin{thebibliography}{51}
\expandafter\ifx\csname natexlab\endcsname\relax\def\natexlab#1{#1}\fi

\bibitem[{Ahn {et~al.}(2012)Ahn, Alexandroff, {Allende Prieto}, Anderson,
  Anderton, Andrews, Aubourg, Bailey, Balbinot, Barnes, Bautista, Beers,
  Beifiori, Berlind, Bhardwaj, Bizyaev, Blake, Blanton, Blomqvist, Bochanski,
  Bolton, Borde, Bovy, Brandt, Brinkmann, Brown, Brownstein, Bundy, Busca,
  Carithers, Carnero, Carr, Casetti-Dinescu, Chen, Chiappini, Comparat,
  Connolly, Crepp, Cristiani, Croft, Cuesta, da~Costa, Davenport, Dawson,
  de~Putter, {De Lee}, Delubac, Dhital, Ealet, Ebelke, Edmondson, Eisenstein,
  Escoffier, Esposito, Evans, Fan, {Femen\'{\i}a Castell\'{a}}, {Fern\'{a}ndez
  Alvar}, Ferreira, {Filiz Ak}, Finley, Fleming, Font-Ribera, Frinchaboy,
  Garc\'{\i}a-Hern\'{a}ndez, P\'{e}rez, Ge, G\'{e}nova-Santos, Gillespie,
  Girardi, {Gonz\'{a}lez Hern\'{a}ndez}, Grebel, Gunn, Guo, Haggard, Hamilton,
  Harris, Hawley, Hearty, Ho, Hogg, Holtzman, Honscheid, Huehnerhoff, Ivans,
  Ivezi\'{c}, Jacobson, Jiang, Johansson, Johnson, Kauffmann, Kirkby,
  Kirkpatrick, Klaene, Knapp, Kneib, {Le Goff}, Leauthaud, Lee, Lee, Long,
  Loomis, Lucatello, Lundgren, Lupton, Ma, Ma, MacDonald, Mack, Mahadevan,
  Maia, Majewski, Makler, Malanushenko, Malanushenko, Manchado, Mandelbaum,
  Manera, Maraston, Margala, Martell, McBride, McGreer, McMahon, M\'{e}nard,
  Meszaros, Miralda-Escud\'{e}, Montero-Dorta, Montesano, Morrison, Muna, Munn,
  Murayama, Myers, Neto, Nguyen, Nichol, Nidever, Noterdaeme, Nuza, Ogando,
  Olmstead, Oravetz, Owen, Padmanabhan, Palanque-Delabrouille, Pan, Parejko,
  Parihar, P\^{a}ris, Pattarakijwanich, Pepper, Percival, P\'{e}rez-Fournon,
  P\'{e}rez-R\`{a}fols, Petitjean, Pforr, Pieri, Pinsonneault, {Porto de
  Mello}, Prada, Price-Whelan, Raddick, Rebolo, Rich, Richards, Robin,
  Rocha-Pinto, Rockosi, Roe, Ross, Ross, Rossi, Rubi\~{n}o Martin, Samushia,
  {Sanchez Almeida}, S\'{a}nchez, Santiago, Sayres, Schlegel, Schlesinger,
  Schmidt, Schneider, Schultheis, Schwope, Sc\'{o}ccola, Seljak, Sheldon, Shen,
  Shu, Simmerer, Simmons, Skibba, Skrutskie, Slosar, Sobreira, Sobeck, Stassun,
  Steele, Steinmetz, Strauss, Streblyanska, Suzuki, Swanson, Tal, Thakar,
  Thomas, Thompson, Tinker, Tojeiro, Tremonti, {Vargas Maga\~{n}a}, Verde,
  Viel, Vikas, Vogt, Wake, Wang, Weaver, Weinberg, Weiner, West, White, Wilson,
  Wisniewski, Wood-Vasey, Yanny, Y\`{e}che, York, Zamora, Zasowski, Zehavi,
  Zhao, Zheng, Zhu, \& Zinn}]{Ahn2012}
Ahn, C.~P., Alexandroff, R., {Allende Prieto}, C., {et~al.} 2012, ApJS, 203, 21

\bibitem[{{Ammons} {et~al.}(2014){Ammons}, {Wong}, {Zabludoff}, \&
  {Keeton}}]{S.MarkAmmonsKennethC.WongCharlesR.Keeton2013}
{Ammons}, S.~M., {Wong}, K.~C., {Zabludoff}, A.~I., \& {Keeton}, C.~R. 2014,
  \apj, 781, 2

\bibitem[{Angulo {et~al.}(2012)Angulo, Springel, White, Jenkins, Baugh, \&
  Frenk}]{Angulo2012}
Angulo, R.~E., Springel, V., White, S. D.~M., {et~al.} 2012, MNRAS, 426, 2046

\bibitem[{Angulo \& White(2010)}]{Angulo2010}
Angulo, R.~E., \& White, S. D.~M. 2010, MNRAS, 405, 143

\bibitem[{Bartelmann {et~al.}(1998)Bartelmann, Huss, Colberg, Jenkins, \&
  Pearce}]{BartelmannMatthias1998}
Bartelmann, M., Huss, A., Colberg, J., Jenkins, A., \& Pearce, F. 1998,
  Astronomy and Astrophysics

\bibitem[{Bouwens {et~al.}(2009)Bouwens, Illingworth, Bradley, Ford, Franx,
  Zheng, Broadhurst, Coe, \& Jee}]{Bouwens2009}
Bouwens, R.~J., Illingworth, G.~D., Bradley, L.~D., {et~al.} 2009, ApJ, 690,
  1764

\bibitem[{Bouwens {et~al.}(2011{\natexlab{a}})Bouwens, Illingworth, Labbe,
  Oesch, Trenti, Carollo, van Dokkum, Franx, Stiavelli, Gonz\'{a}lez, Magee, \&
  Bradley}]{Bouwens2011}
Bouwens, R.~J., Illingworth, G.~D., Labbe, I., {et~al.} 2011{\natexlab{a}},
  Nature, 469, 504

\bibitem[{Bouwens {et~al.}(2011{\natexlab{b}})Bouwens, Illingworth, Oesch,
  Labb\'{e}, Trenti, van Dokkum, Franx, Stiavelli, Carollo, Magee, \&
  Gonzalez}]{Bouwens2011a}
Bouwens, R.~J., Illingworth, G.~D., Oesch, P.~A., {et~al.} 2011{\natexlab{b}},
  ApJ, 737, 90

\bibitem[{Bouwens {et~al.}(2012)Bouwens, Illingworth, Oesch, Trenti, Labb\'{e},
  Franx, Stiavelli, Carollo, van Dokkum, \& Magee}]{Bouwens2012}
---. 2012, ApJ, 752, L5

\bibitem[{Coe {et~al.}(2012)Coe, Umetsu, Zitrin, Donahue, Medezinski, Postman,
  Carrasco, Anguita, Geller, Rines, Diaferio, Kurtz, Bradley, Koekemoer, Zheng,
  Nonino, Molino, Mahdavi, Lemze, Infante, Ogaz, Melchior, Host, Ford, Grillo,
  Rosati, Jim\'{e}nez-Teja, Moustakas, Broadhurst, Ascaso, Lahav, Bartelmann,
  Ben\'{\i}tez, Bouwens, Graur, Graves, Jha, Jouvel, Kelson, Moustakas, Maoz,
  Meneghetti, Merten, Riess, Rodney, \& Seitz}]{Coe2012}
Coe, D., Umetsu, K., Zitrin, A., {et~al.} 2012, ApJ, 757, 22

\bibitem[{{Coe} {et~al.}(2013){Coe}, {Zitrin}, {Carrasco}, {Shu}, {Zheng},
  {Postman}, {Bradley}, {Koekemoer}, {Bouwens}, {Broadhurst}, {Monna}, {Host},
  {Moustakas}, {Ford}, {Moustakas}, {van der Wel}, {Donahue}, {Rodney},
  {Ben{\'{\i}}tez}, {Jouvel}, {Seitz}, {Kelson}, \& {Rosati}}]{Coe2012a}
{Coe}, D., {Zitrin}, A., {Carrasco}, M., {et~al.} 2013, \apj, 762, 32

\bibitem[{Davis {et~al.}(1985)Davis, Efstathiou, Frenk, \& White}]{Davis1985}
Davis, M., Efstathiou, G., Frenk, C.~S., \& White, S. D.~M. 1985, ApJ, 292, 371

\bibitem[{{Ebeling} {et~al.}(2010){Ebeling}, {Edge}, {Mantz}, {Barrett},
  {Henry}, {Ma}, \& {van Speybroeck}}]{Ebeling2010}
{Ebeling}, H., {Edge}, A.~C., {Mantz}, A., {et~al.} 2010, \mnras, 407, 83

\bibitem[{{Ellis} {et~al.}(2013){Ellis}, {McLure}, {Dunlop}, {Robertson},
  {Ono}, {Schenker}, {Koekemoer}, {Bowler}, {Ouchi}, {Rogers}, {Curtis-Lake},
  {Schneider}, {Charlot}, {Stark}, {Furlanetto}, \& {Cirasuolo}}]{Ellis2012}
{Ellis}, R.~S., {McLure}, R.~J., {Dunlop}, J.~S., {et~al.} 2013, \apjl, 763, L7

\bibitem[{Hennawi {et~al.}(2007)Hennawi, Dalal, Bode, \&
  Ostriker}]{Hennawi2007}
Hennawi, J.~F., Dalal, N., Bode, P., \& Ostriker, J.~P. 2007, ApJ, 654, 714

\bibitem[{Hennawi {et~al.}(2008)Hennawi, Gladders, Oguri, Dalal, Koester,
  Natarajan, Strauss, Inada, Kayo, Lin, Lampeitl, Annis, Bahcall, \&
  Schneider}]{Hennawi2008}
Hennawi, J.~F., Gladders, M.~D., Oguri, M., {et~al.} 2008, The Astronomical
  Journal, 135, 664

\bibitem[{{Hilbert} {et~al.}(2007){Hilbert}, {White}, {Hartlap}, \&
  {Schneider}}]{Hilbert2007}
{Hilbert}, S., {White}, S.~D.~M., {Hartlap}, J., \& {Schneider}, P. 2007,
  \mnras, 382, 121

\bibitem[{Ho {et~al.}(2009)Ho, Lin, Spergel, \& Hirata}]{Ho2009}
Ho, S., Lin, Y.-T., Spergel, D., \& Hirata, C.~M. 2009, ApJ, 697, 1358

\bibitem[{Host(2012)}]{Host2012}
Host, O. 2012, MNRAS: Letters, 420, L18

\bibitem[{Ivezic {et~al.}(2008)Ivezic, Tyson, Acosta, Allsman, Anderson,
  Andrew, Angel, Axelrod, Barr, Becker, Becla, Beldica, Blandford, Bloom,
  Borne, Brandt, Brown, Bullock, Burke, Chandrasekharan, Chesley, Claver,
  Connolly, Cook, Cooray, Covey, Cribbs, Cutri, Daues, Delgado, Ferguson,
  Gawiser, Geary, Gee, Geha, Gibson, Gilmore, Gressler, Hogan, Huffer, Jacoby,
  Jain, Jernigan, Jones, Juric, Kahn, Kalirai, Kantor, Kessler, Kirkby, Knox,
  Krabbendam, Krughoff, Kulkarni, Lambert, Levine, Liang, Lim, Lupton,
  Marshall, Marshall, May, Miller, Mills, Monet, Neill, Nordby, O'Connor,
  Oliver, Olivier, Olsen, Owen, Peterson, Petry, Pierfederici, Pietrowicz,
  Pike, Pinto, Plante, Radeka, Rasmussen, Ridgway, Rosing, Saha, Schalk,
  Schindler, Schneider, Schumacher, Sebag, Seppala, Shipsey, Silvestri, Smith,
  Smith, Strauss, Stubbs, Sweeney, Szalay, Thaler, Berk, Walkowicz, Warner,
  Willman, Wittman, Wolff, Wood-Vasey, Yoachim, Zhan, \&
  Collaboration}]{Ivezic2008}
Ivezic, Z., Tyson, J.~A., Acosta, E., {et~al.} 2008, arXiv:0805.2366, 34

\bibitem[{Jullo {et~al.}(2010)Jullo, Natarajan, Kneib, D'Aloisio, Limousin,
  Richard, \& Schimd}]{Jullo2010}
Jullo, E., Natarajan, P., Kneib, J.-P., {et~al.} 2010, Science (New York,
  N.Y.), 329, 924

\bibitem[{Keeton(2001)}]{Keeton2001}
Keeton, C.~R. 2001, arXiv:astro-ph/0102340

\bibitem[{Kitzbichler \& White(2007)}]{Kitzbichler2007}
Kitzbichler, M.~G., \& White, S. D.~M. 2007, MNRAS, 376, 2

\bibitem[{{Kneib} {et~al.}(2003){Kneib}, {Hudelot}, {Ellis}, {Treu}, {Smith},
  {Marshall}, {Czoske}, {Smail}, \& {Natarajan}}]{Kneib2003}
{Kneib}, J.-P., {Hudelot}, P., {Ellis}, R.~S., {et~al.} 2003, \apj, 598, 804

\bibitem[{Komatsu {et~al.}(2011)Komatsu, Smith, Dunkley, Bennett, Gold,
  Hinshaw, Jarosik, Larson, Nolta, Page, Spergel, Halpern, Hill, Kogut, Limon,
  Meyer, Odegard, Tucker, Weiland, Wollack, \& Wright}]{Komatsu2011}
Komatsu, E., Smith, K.~M., Dunkley, J., {et~al.} 2011, ApJS, 192, 18

\bibitem[{Lemze {et~al.}(2009)Lemze, Broadhurst, Rephaeli, Barkana, \&
  Umetsu}]{Lemze2009}
Lemze, D., Broadhurst, T., Rephaeli, Y., Barkana, R., \& Umetsu, K. 2009, ApJ,
  701, 1336

\bibitem[{Maizy {et~al.}(2010)Maizy, Richard, {De Leo}, Pell\'{o}, \&
  Kneib}]{Maizy2009}
Maizy, A., Richard, J., {De Leo}, M.~A., Pell\'{o}, R., \& Kneib, J.~P. 2010,
  Astronomy and Astrophysics, 509, A105

\bibitem[{{Navarro} {et~al.}(1996){Navarro}, {Frenk}, \& {White}}]{Navarro1996}
{Navarro}, J.~F., {Frenk}, C.~S., \& {White}, S.~D.~M. 1996, \apj, 462, 563

\bibitem[{{Oesch} {et~al.}(2012){Oesch}, {Bouwens}, {Illingworth}, {Gonzalez},
  {Trenti}, {van Dokkum}, {Franx}, {Labb{\'e}}, {Carollo}, \&
  {Magee}}]{Oesch2012}
{Oesch}, P.~A., {Bouwens}, R.~J., {Illingworth}, G.~D., {et~al.} 2012, \apj,
  759, 135

\bibitem[{{Oesch} {et~al.}(2013){Oesch}, {Bouwens}, {Illingworth}, {Labb{\'e}},
  {Franx}, {van Dokkum}, {Trenti}, {Stiavelli}, {Gonzalez}, \&
  {Magee}}]{Oesch2013}
---. 2013, \apj, 773, 75

\bibitem[{{Ono} {et~al.}(2013){Ono}, {Ouchi}, {Curtis-Lake}, {Schenker},
  {Ellis}, {McLure}, {Dunlop}, {Robertson}, {Koekemoer}, {Bowler}, {Rogers},
  {Schneider}, {Charlot}, {Stark}, {Shimasaku}, {Furlanetto}, \&
  {Cirasuolo}}]{Ono2012}
{Ono}, Y., {Ouchi}, M., {Curtis-Lake}, E., {et~al.} 2013, \apj, 777, 155

\bibitem[{{Planck Collaboration} {et~al.}(2013){Planck Collaboration}, {Ade},
  {Aghanim}, {Armitage-Caplan}, {Arnaud}, {Ashdown}, {Atrio-Barandela},
  {Aumont}, {Baccigalupi}, {Banday}, \& et~al.}]{PlanckCollaboration2013}
{Planck Collaboration}, {Ade}, P.~A.~R., {Aghanim}, N., {et~al.} 2013,
  arXiv:1303.5076

\bibitem[{Postman {et~al.}(2012)Postman, Coe, Ben\'{\i}tez, Bradley,
  Broadhurst, Donahue, Ford, Graur, Graves, Jouvel, Koekemoer, Lemze,
  Medezinski, Molino, Moustakas, Ogaz, Riess, Rodney, Rosati, Umetsu, Zheng,
  Zitrin, Bartelmann, Bouwens, Czakon, Golwala, Host, Infante, Jha,
  Jimenez-Teja, Kelson, Lahav, Lazkoz, Maoz, McCully, Melchior, Meneghetti,
  Merten, Moustakas, Nonino, Patel, Reg\"{o}s, Sayers, Seitz, \& {Van der
  Wel}}]{Postman2012}
Postman, M., Coe, D., Ben\'{\i}tez, N., {et~al.} 2012, ApJS, 199, 25

\bibitem[{Puchwein \& Hilbert(2009)}]{Puchwein2009}
Puchwein, E., \& Hilbert, S. 2009, MNRAS, 398, 1298

\bibitem[{Richards {et~al.}(2011)Richards, Starr, Butler, Bloom, Brewer,
  Crellin-Quick, Higgins, Kennedy, \& Rischard}]{Richards2011}
Richards, J.~W., Starr, D.~L., Butler, N.~R., {et~al.} 2011, ApJ, 733, 10

\bibitem[{{Robertson} {et~al.}(2013){Robertson}, {Furlanetto}, {Schneider},
  {Charlot}, {Ellis}, {Stark}, {McLure}, {Dunlop}, {Koekemoer}, {Schenker},
  {Ouchi}, {Ono}, {Curtis-Lake}, {Rogers}, {Bowler}, \&
  {Cirasuolo}}]{Robertson2013}
{Robertson}, B.~E., {Furlanetto}, S.~R., {Schneider}, E., {et~al.} 2013, \apj,
  768, 71

\bibitem[{{Ruiz} {et~al.}(2011){Ruiz}, {Padilla}, {Dom{\'{\i}}nguez}, \&
  {Cora}}]{Ruiz2011}
{Ruiz}, A.~N., {Padilla}, N.~D., {Dom{\'{\i}}nguez}, M.~J., \& {Cora}, S.~A.
  2011, \mnras, 418, 2422

\bibitem[{{Shu} {et~al.}(2008){Shu}, {Zhou}, {Bartelmann}, {Comerford},
  {Huang}, \& {Mellier}}]{Shu2008}
{Shu}, C., {Zhou}, B., {Bartelmann}, M., {et~al.} 2008, \apj, 685, 70

\bibitem[{{Spergel} {et~al.}(2003){Spergel}, {Verde}, {Peiris}, {Komatsu},
  {Nolta}, {Bennett}, {Halpern}, {Hinshaw}, {Jarosik}, {Kogut}, {Limon},
  {Meyer}, {Page}, {Tucker}, {Weiland}, {Wollack}, \& {Wright}}]{Spergel2001}
{Spergel}, D.~N., {Verde}, L., {Peiris}, H.~V., {et~al.} 2003, \apjs, 148, 175

\bibitem[{Springel {et~al.}(2001)Springel, Yoshida, \& White}]{Springel2001}
Springel, V., Yoshida, N., \& White, S.~D. 2001, New Astronomy, 6, 79

\bibitem[{Springel {et~al.}(2005)Springel, White, Jenkins, Frenk, Yoshida, Gao,
  Navarro, Thacker, Croton, Helly, Peacock, Cole, Thomas, Couchman, Evrard,
  Colberg, \& Pearce}]{Springel2005}
Springel, V., White, S. D.~M., Jenkins, A., {et~al.} 2005, Nature, 435, 629

\bibitem[{{Sunyaev} \& {Zeldovich}(1980)}]{Sunyaev1980}
{Sunyaev}, R.~A., \& {Zeldovich}, I.~B. 1980, \araa, 18, 537

\bibitem[{Umetsu {et~al.}(2012)Umetsu, Medezinski, Nonino, Merten, Zitrin,
  Molino, Grillo, Carrasco, Donahue, Mahdavi, Coe, Postman, Koekemoer, Czakon,
  Sayers, Mroczkowski, Golwala, Koch, Lin, Molnar, Rosati, Balestra, Mercurio,
  Scodeggio, Biviano, Anguita, Infante, Seidel, Sendra, Jouvel, Host, Lemze,
  Broadhurst, Meneghetti, Moustakas, Bartelmann, Ben\'{\i}tez, Bouwens,
  Bradley, Ford, Jim\'{e}nez-Teja, Kelson, Lahav, Melchior, Moustakas, Ogaz,
  Seitz, \& Zheng}]{Umetsu2012}
Umetsu, K., Medezinski, E., Nonino, M., {et~al.} 2012, ApJ, 755, 56

\bibitem[{Wambsganss {et~al.}(2005)Wambsganss, Bode, \&
  Ostriker}]{Wambsganss2005}
Wambsganss, J., Bode, P., \& Ostriker, J.~P. 2005, ApJ, 635, L1

\bibitem[{Wong {et~al.}(2012)Wong, Ammons, Keeton, \& Zabludoff}]{Wong2012}
Wong, K.~C., Ammons, S.~M., Keeton, C.~R., \& Zabludoff, A.~I. 2012, ApJ, 752,
  104

\bibitem[{{Wong} {et~al.}(2013){Wong}, {Zabludoff}, {Ammons}, {Keeton}, {Hogg},
  \& {Gonzalez}}]{Wong2013}
{Wong}, K.~C., {Zabludoff}, A.~I., {Ammons}, S.~M., {et~al.} 2013, \apj, 769,
  52

\bibitem[{Zhao {et~al.}(2009)Zhao, Jing, Mo, \& B\"{o}rner}]{Zhao2009}
Zhao, D.~H., Jing, Y.~P., Mo, H.~J., \& B\"{o}rner, G. 2009, ApJ, 707, 354

\bibitem[{Zheng {et~al.}(2012)Zheng, Postman, Zitrin, Moustakas, Shu, Jouvel,
  H\o~st, Molino, Bradley, Coe, Moustakas, Carrasco, Ford, Ben\'{\i}tez, Lauer,
  Seitz, Bouwens, Koekemoer, Medezinski, Bartelmann, Broadhurst, Donahue,
  Grillo, Infante, Jha, Kelson, Lahav, Lemze, Melchior, Meneghetti, Merten,
  Nonino, Ogaz, Rosati, Umetsu, \& van~der Wel}]{Zheng2012}
Zheng, W., Postman, M., Zitrin, A., {et~al.} 2012, Nature, 489, 406

\bibitem[{{Zheng} {et~al.}(2009){Zheng}, {Zehavi}, {Eisenstein}, {Weinberg}, \&
  {Jing}}]{Zheng2009b}
{Zheng}, Z., {Zehavi}, I., {Eisenstein}, D.~J., {Weinberg}, D.~H., \& {Jing},
  Y.~P. 2009, \apj, 707, 554

\bibitem[{Zitrin {et~al.}(2011)Zitrin, Broadhurst, Coe, Umetsu, Postman,
  Ben\'{\i}tez, Meneghetti, Medezinski, Jouvel, Bradley, Koekemoer, Zheng,
  Ford, Merten, Kelson, Lahav, Lemze, Molino, Nonino, Donahue, Rosati, {Van der
  Wel}, Bartelmann, Bouwens, Graur, Graves, Host, Infante, Jha, Jimenez-Teja,
  Lazkoz, Maoz, McCully, Melchior, Moustakas, Ogaz, Patel, Regoes, Riess,
  Rodney, \& Seitz}]{Zitrin2011a}
Zitrin, A., Broadhurst, T., Coe, D., {et~al.} 2011, ApJ, 742, 117

\bibitem[{{Zitrin} {et~al.}(2013){Zitrin}, {Meneghetti}, {Umetsu},
  {Broadhurst}, {Bartelmann}, {Bouwens}, {Bradley}, {Carrasco}, {Coe}, {Ford},
  {Kelson}, {Koekemoer}, {Medezinski}, {Moustakas}, {Moustakas}, {Nonino},
  {Postman}, {Rosati}, {Seidel}, {Seitz}, {Sendra}, {Shu}, {Vega}, \&
  {Zheng}}]{Zitrin2013}
{Zitrin}, A., {Meneghetti}, M., {Umetsu}, K., {et~al.} 2013, \apjl, 762, L30

\end{thebibliography}

\clearpage

\end{document}